\documentclass[11pt, a4paper]{article}
\pdfoutput=1

\usepackage{jcappub}

\usepackage{comment}
\usepackage{amsmath}
\usepackage{amsfonts}
\usepackage{amssymb}
\usepackage{graphicx}
\usepackage{mathrsfs}
\usepackage{latexsym}
\usepackage{color}
\usepackage{slashed, cancel}
\usepackage{hyperref}
\usepackage{cleveref}
\usepackage{float}
\usepackage{tikz}
\usepackage{bbold}
\usepackage{soul} 
\usepackage{mathtools}
\usepackage{support-caption} 
\usepackage{subcaption}
\usepackage{enumitem}
\usepackage{journals}
\allowdisplaybreaks

\hypersetup{urlcolor=blue, colorlinks=true} 

\usepackage{fancyhdr}
\numberwithin{equation}{section}

\definecolor{rossos}{rgb}{0.8,0.2,0.3}
\definecolor{bluscuro}{rgb}{0.15, 0.2, .85}
\definecolor{bluchiaro}{cmyk}{1,.3,0.,0.1}
\hypersetup{colorlinks, citecolor=bluscuro, linkcolor=bluscuro, urlcolor=bluscuro}

\makeatother   
\newcommand{\GeV}{{\rm \,GeV}}
\newcommand{\TeV}{{\rm \,TeV}}

\newcommand{\mathsc}[1]{\text{\textsc{#1}}}

\newcommand{\met}{\slashed{E}_T}

 \def\be   {\begin{equation}}   \def\ee   {\end{equation}}
 \def\ba   {\begin{array}}      \def\ea   {\end{array}}
 \def\bea  {\begin{eqnarray}}   \def\eea  {\end{eqnarray}}
 \def\bean {\begin{eqnarray*}}  \def\eean {\end{eqnarray*}}
 \def\nn{\nonumber}

\begin{document}

\today

\title{Two Higgs Doublet Dark Matter Portal}

\author{Nicole F.\ Bell,}
\author{Giorgio Busoni and}
\author{Isaac W. Sanderson}
\affiliation{ARC Centre of Excellence for Particle Physics at the Terascale \\
School of Physics, The University of Melbourne, Victoria 3010, Australia}

\emailAdd{\tt n.bell@unimelb.edu.au}
\emailAdd{\tt giorgio.busoni@unimelb.edu.au}
\emailAdd{\tt isanderson@student.unimelb.edu.au}

\abstract{ We study a fermionic dark matter model in which the
  interaction of the dark and visible sectors is mediated by Higgs
  portal type couplings.  Specifically, we consider the mixing of a
  dark sector scalar with the scalars of a Two Higgs Doublet Model
  extension of the Standard Model. Given that scalar exchange will
  result in a spin-independent dark matter-nucleon scattering cross section,
  such a model is potentially subject to stringent direct detection
  constraints.  Moreover, the addition of new charged scalars
  introduce non-trivial flavour constraints. Nonetheless, this model
  allows more freedom than a standard Higgs portal scenario involving
  a single Higgs doublet, and much of the interesting parameter space
  is not well approximated by a Simplified Model with a single scalar
  mediator. We perform a detailed parameter scan to determine the
  mass and coupling parameters which satisfy direct
  detection, flavour, precision electroweak, stability, and perturbativity constraints, while still producing the correct relic density through thermal freezeout.
 }
\maketitle


\section{Introduction}

The Standard Model of Particle Physics (SM) has seen incredible
success in explaining the physical world, as well as predicting new
particles -- the most recent being the discovery of the 125 GeV Higgs
Boson~\citep{Chatrchyan:2012xdj,Aad:2012tfa}. The mounting
cosmological and astrophysical evidence for Dark Matter (DM), however,
provides a persistent reminder that an extension of this theory is
necessary.
Many existing SM extensions provide viable DM candidates --
supersymmetry, axions, and sterile neutrinos to name a
few. One general class of DM candidates are Weakly Interacting Massive
Particles (WIMPs). These are favoured phenomenologically due to the so-called ``WIMP miracle", in which a particle with mass and coupling
parameters comparable to those of the electroweak interactions can
reproduce the correct relic abundance via thermal
freezeout~\citep{Gondolo:1990dk,Bertone:2004pz}.

To interpret searches for DM signals, it is necessary to somehow model the
details of the particle interactions.  Since there are such a
large number of theories that provide a possible DM candidate,
model-independent descriptions that capture the pertinent features of
the physics are desirable. The most simplistic way to model the
interaction, while still allowing all possible Lorentz structures, is
to construct an Effective Field Theory (EFT).  The interactions are
then described by non-renormalizable operators, obtained from some
more complete theory (about which we remain agnostic) by integrating
out higher energy degrees of freedom.  Such a description will be
valid provided that the mass scale of the mediating particle(s) is
much larger than the relevant momentum transfer.  For example, EFTs
are typically a good approximation for Direct Detection (DD), as the
momentum transfer in DM-nucleon scattering is sub-GeV and hence
significantly smaller than the electroweak scale at which new DM
physics might be expected to emerge.

EFTs, however, suffer potential validity issues at collider
energies~\citep{Busoni:2013lha, Busoni:2014sya, Busoni:2014haa,
  Buchmueller:2013dya, Shoemaker:2011vi, Abdallah:2014hon,Bell:2016obu} where the
momentum transfer may be comparable to or larger than the mass of the
particles that have been integrated out.  For this reason, DM EFTs
were quickly superseded by the Simplified Model
framework~\citep{Abdallah:2014hon,Abdallah:2015ter} which, in addition
to the DM candidate, explicitly introduces a particle to mediate the
interactions.  This is clearly superior to the EFTs, as it enables
resonances or on-shell production of mediators to be considered.  The
three most commonly considered Simplified Models involve SM fermions
interacting with fermionic DM via the exchange of a gauge singlet
spin-0 or spin-1 mediator in the s-channel, or a colored spin-0
mediator in the
t-channel~\citep{Abercrombie:2015wmb,Boveia:2016mrp,Jacques:2016dqz,Buckley:2014fba,Harris:2014hga}.

While the Simplified Models provide a convenient form to constrain,
the earliest versions studied were often too simple to be physically
self-consistent. This is because they did not impose desirable
constraints such as gauge invariance, unitarity, and
renormalizability~\citep{Bell:2015sza, Bell:2015rdw, Haisch:2016usn,
  Englert:2016joy, Kahlhoefer:2015bea, Bell:2016fqf, Ko:2016zxg,
  Duerr:2016tmh, Bell:2016uhg}; ad hoc field additions are required to
restore these features.  For example, if a massive spin-1 mediator has
non-zero axial vector couplings, new physics is required to avoid
violating perturbative unitarity. In this case, if we assume the
spin-1 field is a gauge boson of some dark-sector symmetry, unitarity
can be satisfied via the introduction of a dark Higgs to unitarize the
longitudinal polarization of the
mediator~\citep{Kahlhoefer:2015bea}. This can meaningfully alter the
phenomenology, and open possible new discovery
channels~\citep{Bell:2016fqf,Bell:2016uhg}.
Moreover, even invariance under SM gauge symmetries was not originally
imposed on the Simplified Models.  An example is a gauge singlet
spin-1 mediator which couples with unequal strength to left-handed up and down
type quarks, thus breaking weak isospin. This can lead to the
cross sections for certain processes, such the mono-$W$ collider
signal, increasing indefinitely with the center of mass energy
\citep{Bell:2015sza, Bell:2015rdw, Haisch:2016usn}; something that is
not possible in a UV-complete theory.

Our focus in this paper is the s-channel scalar mediator scenario.
Here it is very clear that a one-mediator simplified model is
inadequate.  If we assume a fermionic DM candidate that is a singlet
under the SM gauge group, the DM bilinear $\overline{\chi}\chi$ can
couple only to an SU(2)$_\textsc{l}$ singlet scalar, while the SM bilinear
$\overline{f}f$ can couple only to an SU(2)$_\textsc{l}$ doublet scalar
mediator. Therefore, the minimal gauge invariant model involves the
mixing of a singlet scalar with the SM Higgs doublet, as described in~\citep{Khoze:2015sra,Baek:2015lna,Bauer:2016gys,Robens:2016xkb,Wang:2015cda,Lopez-Val:2014jva,Costa:2015llh,Dupuis:2016fda,Balazs:2016tbi,Ko:2016ybp,Lewis:2017dme,Chen:2014ask,Chen:2017qcz}.
This case is heavily constrained, particularly when the DM is
relatively light, by a combination of Higgs invisible width
measurements~\citep{Aad:2015uga, Chatrchyan:2014tja,
  ATLAS-CONF-2015-044}, direct detection constraints~\citep{Bell:2016ekl}, and precision electroweak
constraints~\citep{Lopez-Val:2014jva,Bell:2016ekl}.

The simplest way to allow greater freedom for scalar mediator models
is to mix the singlet scalar not with the SM Higgs, but with
\emph{another} SU(2)$_\textsc{l}$ doublet scalar.  This can be readily
accomplished using the framework of a Two Higgs Doublet Model (2HDM).
This is a familiar and well-studied extension to the SM
\citep{Branco:2011iw,Amaldi:1991cn,Carena:1995wu,Bhattacharyya:2015nca}
arising predominantly, but not exclusively, in supersymmetric\footnote{The NMSSM already contains an additional singlet scalar, on the top of the additional Higgs doublet, and all the scalars can potentially mix, as in \cite{Badziak:2015exr,Badziak:2017uto}.} and
Grand Unified Theories. A number of recent papers have discussed the
mixing of a gauge singlet pseudoscalar mediator with the pseudoscalar of 
a 2HDM~\citep{Ipek:2014gua,Berlin:2015wwa,Goncalves:2016iyg,No:2015xqa,Bauer:2017ota,Haisch:2016gry,Tunney:2017yfp}.
Given that the SM itself contains no pseudoscalar boson (after
symmetry breaking) this is the minimal gauge invariant model in which
DM interacts with the SM via a pseudoscalar mediator.

This paper will instead consider the mixing of a CP-even scalar
singlet with the additional real neutral scalar of a 2HDM, after
symmetry breaking.  In previous work, we outlined the construction of
such 2HDM+S models, with a focus on possible Yukawa
structures~\citep{Bell:2016ekl}. Our aim in the present paper is to
perform a thorough analysis of the viable parameter space of the 2HDM+S
model, and discuss its associated phenomenology. This will be achieved
by performing a detailed scan over model parameters, to identify
values which satisfy relic density and direct detection
constraints. We shall also impose constraints on the charged scalars
arising from flavour physics, along with stability, perturbativity and
electroweak precision requirements.  Importantly, we shall find a
significant portion of viable parameter space that cannot be
approximated by a naive single-mediator Simplified Model.

The 2HDM+S has a richer spectrum of possible DM annihilation modes
than a single-mediator model.  Besides the obvious additional
annihilation channels due to the extra scalar, final states consisting
of charged scalars, pseudoscalars, and gauge bosons are also present.
These additional channels, such as $\bar{\chi} \chi \to Z^0 A$ (where
$A$ is the pseudoscalar of the additional SU(2)$_\textsc{l}$ doublet)
can contribute significantly to the annihilation cross section.
Furthermore, the DM-nucleon cross sections relevant for direct
detection are sensitive not only to the interference of contributions
from the mixed scalar mediators but also, depending on the Yukawa
structure, to relative cancellations between different quarks in the
nucleus. As such, the additional freedom in the Yukawa sector of the
2HDM allows DM-nucleon scattering to be suppressed without hindering
relic density production.

We outline 2HDM+S model in Section~\ref{sec:model} and describe the
constraints in Section~\ref{sec:constraints}. We discuss the results
of our parameter scan in Section~\ref{sec:relicnumeric} and present
our conclusions in Section~\ref{sec:conclusions}.

\section{The 2HDM+S Gauge Invariant Simplified Model: 2HDM + Singlet Scalar S}
\label{sec:model}
We shall consider a Dirac DM candidate, $\chi$, and expand the scalar
sector of the SM to include two Higgs doublets, $\Phi_1$ and
$\Phi_2$, in addition to a real singlet scalar field, $S$.  
In the following two subsections we outline the scalar potential, which controls the mixing between the CP even scalars, and the Yukawa structure, which dictates the coupling of those scalars to the DM and SM fermions.

\subsection{The Scalar Sector}
It is convenient to rotate $\{\Phi_1,\Phi_2\}$ to the Higgs basis $\{
\Phi_h,\Phi_H \}$, which is defined such that only one of the two
doublets obtains a vev\footnote{We can always rotate to the
Higgs basis, without loss of generality. The transformation from
an arbitrary basis to the Higgs basis is outlined in Appendix
\ref{sec:gaugetohiggs}.}. Taking $\langle \Phi_H \rangle = 0$ and 
$\langle \Phi_h \rangle = v \sim 246$ GeV, the two Higgs doublets 
are then defined by
\bea
\Phi_h &=& \cos\beta \Phi_1 + \sin\beta \Phi_2 = \left(
\begin{array}{cc}
 G^+ \\
\frac{v + h + i G^0}{\sqrt{2}} \\
\end{array}
\right),\label{eq:alignh}\\
\Phi_H &=& -\sin\beta \Phi_1 + \cos\beta \Phi_2 = \left(
\begin{array}{cc}
 H^+ \\
\frac{ H + i A}{\sqrt{2}} \\
\end{array}
\right).
\eea

The scalar potential consists of the usual 2HDM potential, augmented by terms involving the new singlet scalar $S$.
We will assume that the scalar potential has a spontaneously broken $\mathbb{Z}_2$ symmetry for the particle $S$. This may arise naturally, for example, in the case where $S$ is part of a complex scalar charged under a dark $U(1)$ gauge group.  The scalar potential is thus given by
\begin{equation}
\hat{V}(\Phi_h,\Phi_H,S) = \hat{V}_{\mathsc{2hdm}}(\Phi_h,\Phi_H) + \hat{V}_S(S) + \hat{V}_{S\mathsc{2hdm}}(\Phi_h,\Phi_H,S), \label{eq:potential}
\end{equation}
where\nobreak
\bea
\hat{V}_{\mathsc{2hdm}}(\Phi_h,\Phi_H) &=& \hat{M}_{hh}^2 \Phi_h^\dagger \Phi_h + \hat{M}_{HH}^2 \Phi_H^\dagger \Phi_H +  (\hat{M}_{hH}^2 \Phi_H^\dagger \Phi_h + h.c.) + \frac{\hat{\lambda}_h}{2} (\Phi_h^\dagger \Phi_h)^2 + \frac{\hat{\lambda}_H}{2} (\Phi_H^\dagger \Phi_H)^2 \nonumber \\
&+&\hat{\lambda}_3 (\Phi_h^\dagger \Phi_h)(\Phi_H^\dagger \Phi_H) + \hat{\lambda}_4 (\Phi_H^\dagger \Phi_h)(\Phi_h^\dagger \Phi_H)  
+ \frac{\hat{\lambda}_5}{2} \left( (\Phi_H^\dagger \Phi_h)^2 + h.c.\right),\\
\hat{V}_S(S) &=& \frac{1}{2} \hat{M}_{SS}^2 S^2 + \frac{1}{4} \hat{\lambda}_S S^4,
\\
\hat{V}_{S\mathsc{2hdm}}(\Phi_h,\Phi_H,S) &=& 
 \frac{\hat{\lambda}_{HHS}}{2}(\Phi_H^\dagger \Phi_H)S^2 +  \frac{\hat{\lambda}_{hhs}}{2} \Phi_h^\dagger \Phi_h S^2 + \frac{1}{2}(\hat{\lambda}_{hHS} \Phi_H^\dagger \Phi_h S^2 + h.c.).
\eea

In general, there would be mass mixing between all three neutral CP even scalars of the model, $h$, $H$, and $S$. We shall, however, impose a generalized Higgs ``alignment limit" which decouples the SM Higgs, $h$, from the other two states. This is desirable because it reduces the scalar mixing to a 2-state problem and guarantees that $h$ couples like the SM Higgs. We thus set
\begin{align}
  \label{eq:align}  \hat{\lambda}_h &= \hat{\lambda}_H = \hat{\lambda}_3 + \hat{\lambda}_4 + \hat{\lambda}_5, \\
    \hat{\lambda}_{hhs} &= 0,
\end{align}
where the first of these conditions is sufficient to impose alignment in a standard 2HDM, and the second clearly prevents $h$-$S$ mixing (refer to Appendix~\ref{sec:gaugetohiggs} for further discussion of the alignment limit).
The remaining $H$-$S$ mass matrix is then diagonalized to obtain two mass eigenstate scalars, $S_1$ and $S_2$, such that
\begin{align}
H &= \cos\theta S_1 - \sin\theta S_2,\\
S &= v_S + \sin\theta S_1 + \cos\theta S_2,
\end{align}
where 
\begin{equation}
\sin2\theta = \frac{2 \hat{\lambda}_{hHs} v v_S }{M_{S_1}^2-M_{S_2}^2}.
\end{equation}
The scalar mass spectrum then simplifies to
\begin{eqnarray}
\label{eq:mh} M_{h}^2 &=& \hat{\lambda}_h v^2, \\
M_{H^+}^2 &=&  M_{HH}^2+\frac{1}{2} \left(\hat{\lambda}_{HHs}+2\tan\beta \hat{\lambda}_{hHs}\right) v_S^2+\hat{\lambda}_{3}\frac{v^2}{2},\\
M_A^2 &=& M_{H^+}^2 + \left(
  \hat{\lambda}_4-\hat{\lambda}_5 \right)\frac{v^2}{2} ,\\
\label{MS12} M_{S_{1,2}}^2 &=&
\frac{1}{2}\left(M_A^2+\hat{\lambda}_5 v^2\right)\left(1\pm \frac{1}{\cos
  2\theta}\right) + \hat{\lambda}_S v_S^2\left(1\mp \frac{1}{\cos
  2\theta}\right). 
\end{eqnarray}
Taking $h$ to be the observed SM Higgs boson then fixes $\hat{\lambda}_h$ from \eqref{eq:mh} and hence also fixes $\hat{\lambda}_H$ via the alignment condition \eqref{eq:align}. Using the previous equations, we can rewrite $M_A$ and $M_{H^+}$ as a function of $M_{S_{1,2}}$, $\theta$, and $\hat{\lambda}_{4,5}$:
\begin{eqnarray}
M_{H^+}^2 &=&  M_{S_1}^2\cos^2\theta + M_{S_2}^2\sin^2\theta - \left(
  \hat{\lambda}_4+\hat{\lambda}_5 \right)\frac{v^2}{2},\\
M_A^2 &=& M_{S_1}^2\cos^2\theta + M_{S_2}^2\sin^2\theta -\hat{\lambda}_5 v^2.
\end{eqnarray}

\subsection{The Yukawa Sector}

The only portal between the DM and other fields is via its Yukawa
coupling to the singlet scalar,
\begin{align}
\mathcal{L}_\mathsc{dm} &= -y_\chi S \bar{\chi} \chi.
\end{align}
We will assume that the DM particle has no bare mass term, and that its mass is instead generated by the vacuum expectation value of the singlet scalar, i.e. $m_\chi = y_\chi v_s$ with $\langle S \rangle = v_s$. Although this is not strictly necessary, such a scenario arises naturally if DM is a chiral fermion charged under some dark gauge group that is broken spontaneously by the vev of $S$. This assumption adds a constraint between the DM mass and the DM Yukawa coupling, removing the freedom to accommodate the relic density by varying these two parameters independently. 

We will express the Yukawa interactions of the SM fermions with the Higgs doublets as
\begin{equation}
L_{\text{Yukawa}} = - \sum_{n=h,H} \left(Y_{n,ij}^U \bar{Q}_L^i u_R^j \widetilde{\Phi}_n
+ Y_{n,ij}^D \bar{Q}_L^i d_R^j \Phi_n
+ Y_{n,ij}^L \bar{L}_L^i l_R^j \Phi_n + h.c. \right),
\label{eq:yukawah1h2}
\end{equation}
and we will assume that the Yukawa matrices of the additional doublet are proportional to the SM ones: 
\begin{align}
Y_h^i & \equiv Y_{\mathsc{sm}}^i,\\
Y_H^i & = \epsilon_i Y_{\mathsc{sm}}^i,
\end{align}
where the $\epsilon_i$ are Yukawa scaling factors, with $i=u,d,l$. This Yukawa structure is the so-called Aligned Yukawa model \citep{Pich:2009sp,Tuzon:2010vt,Pich:2010ic,Penuelas:2017ikk,Gori:2017qwg}, which satisfies Natural Flavour Conservation. In special cases where the $\epsilon_i$ satisfy certain relationships, the Aligned Yukawa structure can correspond to one of the $Z_2$ symmetric Yukawa structures (Type I, II, X or Y), as shown in Table \ref{tab:coeffs}. While we will determine constraints for both Type I and Type II Yukawa structures, we will also include results for more general choices of the scaling factors that satisfy the Aligned Yukawa criteria. See \citep{Bell:2016ekl} for a more detailed discussion of the Yukawa structure in these models.

\begin{table}[tb]\centering
\begin{tabular}{|c|c|c|c|}
\hline
Model & $\epsilon_d$ & $\epsilon_u$ & $\epsilon_l$
\\ \hline
Type I & $\cot\beta$ & $\cot\beta$ & $\cot\beta$
\\
Type II & $-\tan\beta$ & $\cot\beta$ & $-\tan\beta$
\\
Type X & $\cot\beta$ & $\cot\beta$ & $-\tan\beta$
\\
Type Y & $-\tan\beta$ & $\cot\beta$ & $\cot\beta$
\\
Inert & 0 & 0 & 0
\\ \hline
\end{tabular}
\caption{Values of the Yukawa scaling factors, $\epsilon_{u,d,l}$ which
  correspond to models with discrete ${\cal Z}_2$
  symmetries.}\label{tab:coeffs}
\end{table}

\section{Constraints on the 2HDM+S scenario}
\label{sec:constraints}

In this section we will outline the constraints that apply to our model arising from relic density requirements, DD data, flavour, stability of the potential, and precision electroweak observables.

\subsection{DM Annihilation Processes}
\label{sec:anncs}

\begin{figure}
    \centering
    \hspace{2em} \includegraphics{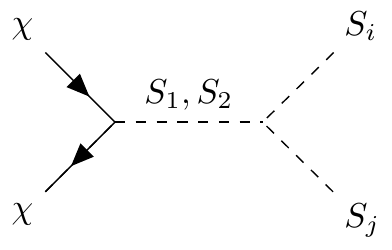} \hspace{3em} \vspace{2em} \includegraphics{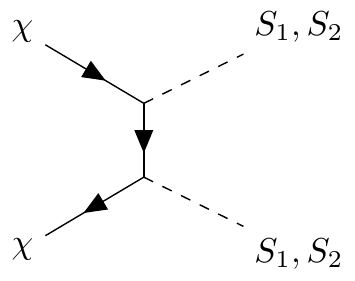} \hspace{3em} \includegraphics{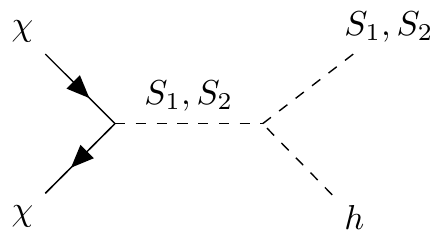}  \hspace{0.5em} \\ \includegraphics{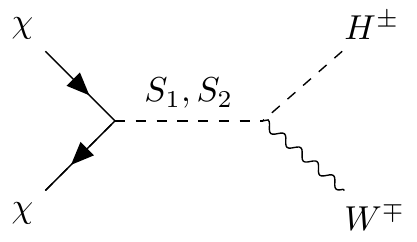} \hspace{2em} \includegraphics{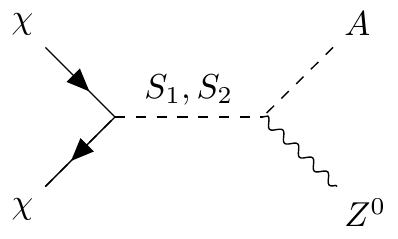} \hspace{2em} \includegraphics{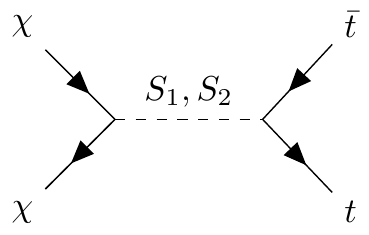} 
    \caption{Dominant DM annihilation channels, where $(S_i,S_j)$ is one of these scalar final states: $(S_1,S_1)$, $ (S_1,S_2)$, $(S_2,S_2)$, $(H^+,H^-)$, $(A,A)$.}
    \label{fig:feyn}
\end{figure}

The 2HDM+S scenario has multiple DM annihilation channels.
There are tree-level annihilations to fermion-antifermion pairs,
scalars (the SM Higgs, the two neutral scalars, the pseudoscalar, and the
charged scalars) and/or the electroweak gauge bosons,
namely: $\overline{\chi}\chi \rightarrow \overline{f}f$, $S_1 S_1$, $S_2
S_2$, $S_1 S_2$, $H^+ H^-$, $H^+ W^-$, $A A$, $A Z$, $S_1 h$, and $S_2
h$ -- these are shown in Fig.~\ref{fig:feyn}. This is to be compared with a single mediator model in which only
the $\overline{f}f$ and $SS$ channels are present. Note that since all
diagrams involve $\chi$-scalar vertices (including those with gauge
boson final states) they are all $p$-wave processes. As such, while we
will easily be able to find parameters that accommodate the observed
relic density, there will be no constraints arising from indirect
detection because the $p$-wave annihilation processes are highly velocity
suppressed in the late universe.

If the DM particle is relatively light, such that annihilation to the
scalars and electroweak bosons is kinematically forbidden ($m_\chi \lesssim
80\GeV$), the only annihilation channels that remain open are the
fermionic ones.  This case is heavily constrained, as the dominant
annihilation channel is then $b\bar{b}$, which is suppressed by the
bottom Yukawa coupling and thus usually requires resonant enhancement
to accommodate the correct relic density.

If, instead, the DM particle is heavy enough to annihilate to the Higgs,
electroweak gauge bosons and/or the new scalars, then these final states will
likely dominate due to the Yukawa suppression of annihilations to fermions 
(top excluded).
Because all of these annihilations are governed by scalar and electroweak
couplings -- and exist due to gauge invariance, independent of the
Yukawa couplings of the second doublet -- they are also present,
for example, in the limit where the second doublet is inert.
This ability to produce the correct relic abundance independent of
Yukawa structure means that DM can be adequately produced while
avoiding any Yukawa dependent constraints (e.g. DD, neutral meson
mixing, $b \to s \gamma$, $\mu \to e \gamma$, etc.).

We implemented the model in Feynrules\footnote{The Feynrules model file used is publicly available in the Feynrules model database.} \citep{Christensen:2008py,Alloul:2013bka} and output the model with the CALCHEP interface~\citep{Christensen:2009jx}. We then used {\tt micrOMEGAs}~\citep{Barducci:2016pcb} to perform the relic density calculation, where we included 3 body final states with off-shell gauge bosons. We also double checked the results by calculating the annihilation cross sections, which are reported in Appendix~\ref{sec:relicdensitycs}. In the case where the parameter values are away from resonances and annihilation thresholds, one can use the wave expansion of the cross sections. The p-wave coefficients of this expansion are also reported in Appendix~\ref{sec:relicdensitycs}. 

Sommerfeld enhancement can significantly increase the DM annihilation
cross section~\citep{Feng:2010zp,Cassel:2009wt,Iengo:2009ni}, provided
at least one of the scalars is both sufficiently light compared to the
DM, and strongly coupled to the DM particle.  In practice, for the
parameter range we consider, this leads to $\mathcal{O}(1)$
corrections to the cross section; a discussion is provided in
Appendix~\ref{sec:somm}.

\subsection{Direct Detection}
\label{sec:dd}

\begin{figure}
    \centering
    \includegraphics{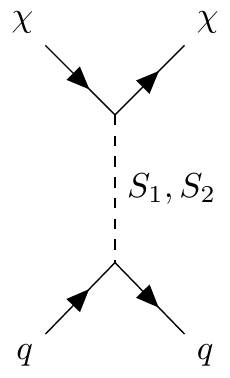} 
    \caption{Spin-independent DM-nucleon scattering arises from the exchange of the mixing scalar mediators.}
    \label{fig:feynDD}
\end{figure}

We will generate DD constraints using the 2016 LUX~\citep{Akerib:2016vxi} and XENON1T~\citep{Aprile:2017iyp} data, via an effective operator approach using tools from \citep{DelNobile:2013sia}. The scattering of DM with nuclei will be dominated by the diagram shown in Fig.~\ref{fig:feynDD}, resulting in a spin-independent scattering cross section. The only relevant nucleon operator is
\begin{align}
O_1^N = \bar{\chi} \chi \bar{N} N,  
\end{align}
and, by integrating out the mediators, we obtain a coefficient of \citep{Bell:2016ekl}
\begin{align}
c_N = m_N\frac{y_\chi \cos\theta\sin\theta}{v}\left(\frac{1}{M_{S_1}^2}-\frac{1}{M_{S_2}^2}\right) \left(f_{T_u}^N\epsilon_u + \epsilon_d\sum_{q=d,s} f_{T_q}^N + \frac{2}{9} f_{T_g} \frac{2 \epsilon_u+ \epsilon_d}{3} \right).
\label{eq:DDcoeff}
\end{align}

As there are contributions from exchange of the two scalars, with a relative negative sign, there is the possibility for destructive interference when the masses of $S_1$ and $S_2$ are comparable. 
Alternatively, it is possible to have destructive interference between the up-type quarks and the down-type quarks in a scattering process. 
While the $S_1$-$S_2$ interference will be a feature of any model with two scalar mediators (such as a singlet scalar mixed with the SM Higgs) the $u$-$d$ interference requires the additional Yukawa freedom provided by the 2HDM framework.
If we adopt the values of $f_{T_i}$ obtained by \citep{Gondolo:2004sc}, then one finds that a ratio of $\epsilon_u \sim - 1.6 \epsilon_d$ will result in exact cancellation of the DD signal\footnote{Note, however, that the values of $f_{T_i}$ vary between different collaborations (\citep{DelNobile:2013sia, Gondolo:2004sc, Ellis:2008hf, Cheng:2012qr, Belanger:2013oya}), and so the ratio required for a complete cancellation depends on which values are adopted.}.  For example, this cancellation will occur in a Type II 2HDM when $\tan\beta=\sqrt{|\frac{\epsilon_d}{\epsilon_u}|}\sim 0.8$, but not in a Type I model.

It is noteworthy that the relic density annihilation processes will be unaffected by this $u$-$d$ interference, as there is no interference between different final state quarks. 
Moreover, DD constraints can be avoided by taking the scalar mixing to be small (either $\theta\rightarrow 0$ or $\theta\rightarrow\pi/2$), while still being able to produce the correct relic density via annihilation to a pair of scalars. We can thus have a scenario where DD is highly suppressed yet relic density constraints are easily satisfied. Finally, while both the DD cross section and relic abundance will depend on $\epsilon_u$, only DD will be sensitive to $\epsilon_d$, as annihilations to $b\bar{b}$ are usually negligible. Because of this, we will present DD constraints for different possible choices of $\epsilon_d$.

\subsection{Flavour Constraints}
\label{sec:flavour}
\begin{figure}
    \centering
    \includegraphics{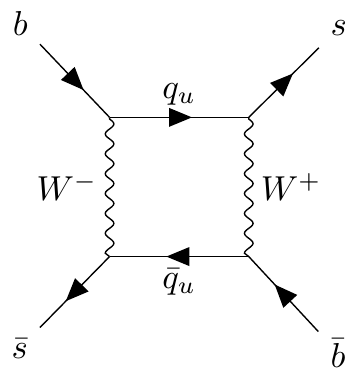}  \includegraphics{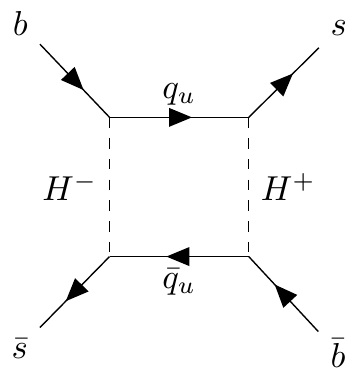}  \includegraphics{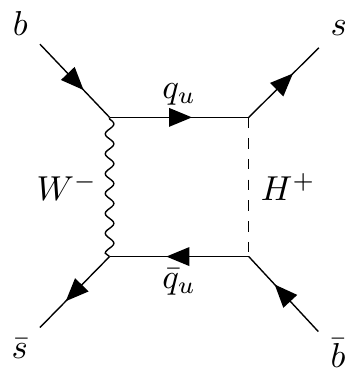}  \includegraphics{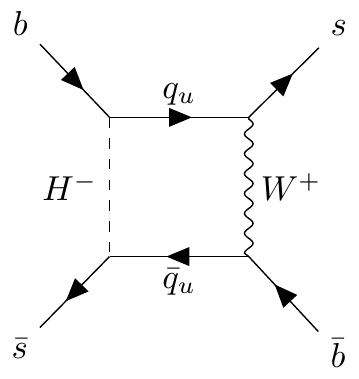}
    \caption{Contribution to the neutral meson mixing, $B_s^0$ -- $\bar{B}_s^0$.  The diagram on the left is the SM contribution, while the three additional diagrams arise in the presence of a charged Higgs.}
    \label{fig:meson}
\end{figure}

As we have chosen to work with an aligned 2HDM, i.e.~where the second Higgs Yukawa couplings are simply proportional to the SM Yukawa couplings, there are no tree-level Flavour Changing Neutral Currents (FCNCs). The presence of a charged Higgs, however, implies flavour changing vertices will still be present, and allows FCNCs to be generated by loops. The strength of the flavour changing processes will therefore depend only on $M_{H^+}$ and on the Yukawa scaling factors, $\epsilon_u$ and $\epsilon_d$, which can then be constrained by flavour physics. For example, the neutral meson mixing $B_s^0$ -- $\bar{B}_s^0$, which is mediated by a $W^\pm$ box diagram in the SM, will receive new contribution from the three diagrams shown in Fig.~\ref{fig:meson} that contain $H^\pm$ exchange. 
Constraints on the mass of the charged Higgs, as a function of the Yukawa scaling factors $\epsilon_u$ and $\epsilon_d$, are taken from \citep{Enomoto:2015wbn}. Note that, at one loop level, our model with the additional singlet is identical to the Aligned 2HDM for the purposes of neutral meson mixing.

\subsection{Electroweak Precision Observables and Stability of the Potential}
\label{sec:ewpt}

We also account for limits on precision electroweak observables. It can be shown that the potential of Eq.~\ref{eq:potential} breaks custodial symmetry \citep{Haber:1992py,Pomarol:1993mu,Barbieri:2006dq,Gerard:2007kn,Grzadkowski:2010dj,Haber:2010bw} (in particular the $\lambda_4,\lambda_5$ and $\lambda_{hHS}$ terms) and leads to contributions to precision electroweak observables, unless $M_A=M_{H^+}$. The most relevant observable is the $\rho$ parameter, which receives a contribution of 
\bea
\Delta\rho = \frac{1}{(4\pi)^2 v^2}  \Big( F(M_{H^+}^2,M_A^2) &+& \cos^2\theta F(M_{H^+}^2,M_{S_1}^2)+\sin^2\theta F(M_{H^+}^2,M_{S_2}^2)\nn\\
&-&\cos^2\theta F(M_A^2,M_{S_1}^2)-\sin^2\theta F(M_A^2,M_{S_2}^2) \Big), 
\label{eq:rhoEW}
\eea
where
\bea
F(x,y) &=& \frac{x+y}{2}-\frac{xy}{x-y}\log(x/y).
\eea
One can verify that $\Delta\rho=0$ for $M_A=M_{H^+}$, while for $M_{S_2}=M_{H^+}$ Eq.~\ref{eq:rhoEW} simplifies to

\bea
\Delta\rho &=& \frac{M_{H^+}^2}{(4\pi)^2 v^2}\left(1+f(M_A^2,M_{S_1}^2,M_{H^+}^2)+f(M_{S_1}^2,M_A^2,M_{H^+}^2)\right)\cos^2\theta,
\eea
where
\bea
f(x,y,z) &=& \frac{x^2(y-z)}{z(x-y)(x-z)} \log(x/z).
\eea
This agrees with the result of \citep{Bauer:2017ota} for the mixed pseudoscalar model (upon replacing
$\cos^2\theta$ with $\sin^2\theta$, due to a difference in the way the mixing angle is defined).

\begin{figure}[t]
    \centering
    \includegraphics[width=0.5\textwidth]{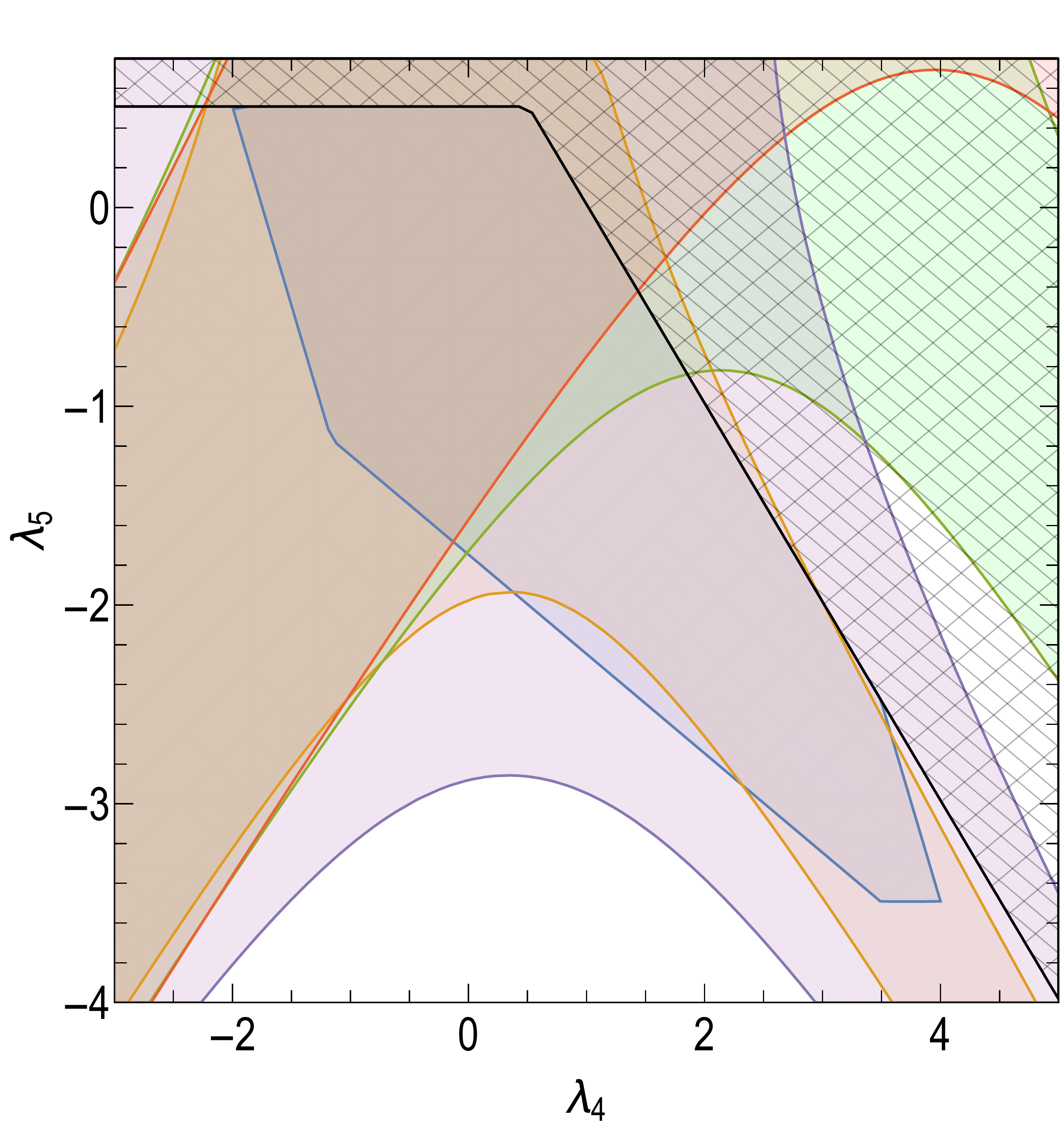} 
    \caption{Parameter space in the $\lambda_4$-$\lambda_5$ plane. The blue region is the one considered in our parameter scan, and satisfies stability and perturbativity conditions. The orange, green, red and purple regions satisfy the $\Delta\rho$ limits for the benchmark points $M_{S_1}=200\GeV$, $M_{S_2}={400,1200,1600,400}\GeV$ and
  $\sin\theta=0.35, 0.35, 0.35, 0.94$,
respectively. The black cross-hatched region is excluded by the bounded from below conditions for the potential at tree level.}
    \label{fig:l4l5}
\end{figure}

Fig.~\ref{fig:l4l5} shows the parameter region we consider for $\lambda_4$, and $\lambda_5$, and some example regions which satisfy the constraint on $\Delta\rho$, for different choices of the scalar masses and mixing parameters. The constraint is significant only when the mixing angle is close to $0$, i.e.~the scalar in the doublet is mostly the lighter of the two, and the mass difference between the two is large.
 
The scalar potential must also satisfy Bounded From Below (BFB) conditions. This was discussed in \citep{Bell:2016ekl} where we derived requirements that are necessary but not sufficient. The BFB conditions alone, however, do not guarantee that the chosen minimum is stable, i.e.~that it is a global minimum. 
Therefore, when performing the parameter scans presented in Section \ref{sec:relicnumeric}, we check that each point chosen lies in a global minimum by numerically minimizing the tree level potential.

Note that this procedure is conservative. Indeed, to check the stability of a point one should calculate the 1 loop corrections to the potential, which can potentially have an important effect on the stability of 2HDM potentials if the scalar couplings are large enough. In particular, Ref.~\citep{Staub:2017ktc} found that most points that are stable at tree level are also stable at 1-loop, while many points that are excluded by the tree level BFB conditions are actually stable when 1-loop corrections are included\footnote{The examples provided in \citep{Staub:2017ktc} use $\lambda_{3,4,5}$ couplings greater than what we allow from perturbativity, and so it is possible that the tree-level constraint is adequate in the majority of the parameter space we scan over.}.
Thus, for simplicity, we chose not to include 1-loop corrections; we expect that including them would result in a small increase in the allowed parameter space.

\subsection{Collider Constraints}
\label{sec:collider}
Dark Matter searches at colliders are currently interpreted in terms of Simplified Models \citep{Abdallah:2014hon,Abdallah:2015ter}. The CMS and ATLAS collaborations have created a set of common interpretation strategies \citep{Abercrombie:2015wmb,Boveia:2016mrp} to  harmonize the way results are interpreted, reported and compared. The first set of Simplified Models \citep{Abdallah:2014hon,Abdallah:2015ter} did not impose gauge invariance, thus the new mediators featured only couplings to DM and quarks. Because of this, the most effective search strategies were Monojet$+\met$ \citep{Khachatryan:2014rra,Aad:2015zva,Sirunyan:2017hci,ATLAS:2017dnw}, Di-jet  \citep{Chatrchyan:2013qha,Khachatryan:2016ecr,Sirunyan:2016iap,Khachatryan:2015dcf,ATLAS:2016xiv,ATLAS:2016bvn} or $b\bar{b}$, $t\bar{t}$ \citep{Aaboud:2016nbq,ATLAS:2016fol,Khachatryan:2015sma,Aad:2015fna} resonance searches and, in the case of scalar mediators, $ b\bar{b}+\met$, $t\bar{t}+\met$ final states \citep{Khachatryan:2015nua,ATLAS:2016tsc,Aad:2014vea,Sirunyan:2017xgm}. Other channels, like $W/Z+\met$, $\gamma+\met$, $h+\met$ were considered \citep{Aad:2014vka,Aad:2013oja,Khachatryan:2016mdm,Aaboud:2016qgg,Sirunyan:2017onm,Aaboud:2016obm,Aad:2015yga} and despite being much cleaner channels with much smaller backgrounds, were usually not competitive with the former ones. By including gauge invariance, Simplified Models with scalar mediators automatically include new couplings to Higgs and gauge bosons \citep{No:2015xqa,Goncalves:2016iyg,Bauer:2017ota,Bell:2016ekl}. Thanks to this,  $W/Z+\met$ and $h+\met$ can now provide competitive limits with $j+\met$, $t\bar{t}+\met$ and Di-jet searches \citep{Sirunyan:2017hnk,Aaboud:2017yqz,Aaboud:2017uak}. 
In particular, it was pointed out that for certain mass spectrums it is possible to achieve resonant production \citep{No:2015xqa,Goncalves:2016iyg,vonBuddenbrock:2016rmr,Bauer:2017ota}, where the heavier of the two new scalars is produced by gluon fusion with a top loop, and then decays to the lighter of the two scalars together with a SM boson, with the lighter scalar decaying invisibly to DM thereafter. This kind of mechanism has been shown to provide an easily detectable mono-$Z/h$ signature when $M_{S_1}\lesssim 370 \GeV$, $1.2\TeV \gtrsim M_{S_2}\gtrsim M_{S_1}+M_{h/Z}$, and $M_{S_1}>2m_\chi$ \citep{Bauer:2017ota}. The upper bound on $M_{S_1}$ is due to the fact that, at larger masses, the new scalar will decay predominantly to top pairs instead of invisibly, reducing the efficiency of the search. Unfortunately, the required mass spectrum for this resonant production is mutually exclusive with the parameters required to achieve the correct relic density, at least in the minimal scenario with just two new scalars. In fact, as we will see, in the low mass regime 
we require at least one of the two scalars to be lighter than $2 m_\chi$ to be able to achieve the right relic density, making the two benchmarks mutually exclusive.
Other possible search strategies proposed also include vector boson fusion signatures \citep{Brooke:2016vlw,Dutta:2017lny} and trying to detect the interference effects in $b\bar{b}$, $t\bar{t}$, $\tau\bar{\tau}$ resonances \citep{Bernreuther:2015fts,Carena:2016npr}.

\section{Parameter Scan}
\label{sec:relicnumeric}

\subsection{Parameter Ranges}

The independent parameters present in the model are
\be
m_\chi, \quad M_{S_1}, \quad M_{S_2}, \quad y_\chi, \quad \hat{\lambda}_4, \quad \hat{\lambda}_5, \quad \hat{\lambda}_{hHS}, \quad \hat{\lambda}_{HHS}, \quad \epsilon_u \quad \textrm{and} \quad \epsilon_d.
\ee
This set of parameters, through the minima condition and the diagonalization relations, together with the additional constraint $m_\chi = y_\chi v_s$, determine all other parameters of the model\footnote{The phase of the DM Yukawa can always be reabsorbed, so one can chose $y_\chi$ and $v_s$ to be both real and positive.}.
The scan is performed in the following range:
\bea
70\GeV < &m_\chi& < 1\TeV,\\
70\GeV < &M_{S_1}& < M_{S_2} < 2 \TeV, \\
0 < &y_\chi& < 2,\\
&|\hat{\lambda}_{hHS}|& < 2,\\
&|\hat{\lambda}_{HHS}|& < 4,\\
0 < &\epsilon_u& < 1,
\eea
while for $\hat{\lambda}_4$ and $\hat{\lambda}_5$ we scan over the region shown in Fig.~\ref{fig:l4l5}. 
These ranges for the couplings were chosen so that most of the points will satisfy unitarity and perturbativity bounds, which was checked via the scalar scattering matrices as in \citep{Bell:2016ekl}.
To achieve the right relic density, we will see that it will be in general necessary to have $m_\chi\gtrsim M_{S_1}$, and the inequality is strictly required for DM masses below the top mass, as otherwise all considered annihilation channels are closed. In this low DM mass region, however, one needs to take into account Higgs invisible constraints \citep{Aad:2015pla,Khachatryan:2016whc}: 2-body decays forbid the region $2m_\chi<m_h$ and $2M_{S_1}<m_h$, while considering 3-body decays as well further pushes up the lower bound on $M_{S_1}$ to nearly $100\GeV$. The Higgs invisible decays constraints can only be avoided in the $\theta\rightarrow\pi/2$ limit, but in such case the model approaches a decoupled dark sector which is phenomenologically uninteresting\footnote{Relic density requirement can be satisfied in the limit of a decoupled dark sector, in which dark matter annihilates to light dark scalars. However, there would be no signals in collider or direct detection experiments. Moreover, indirect detection is prevented by the p-wave nature of annihilation to scalars, even if small couplings to the SM are included.}. Taking into account these considerations, we have chosen in our scan a conservative lower bound for the DM mass and for the lightest scalar of $70\GeV$.
Points are selected if they have a relic density between $0.1 < \Omega h^2 < 0.14$ and satisfy all bounds from flavour, unitarity, perturbativity, tree level vacuum stability, and DD constraints. 

\begin{figure}[t]
\centering
\begin{subfigure}[t]{0.43\textwidth}
\includegraphics[width=\textwidth]{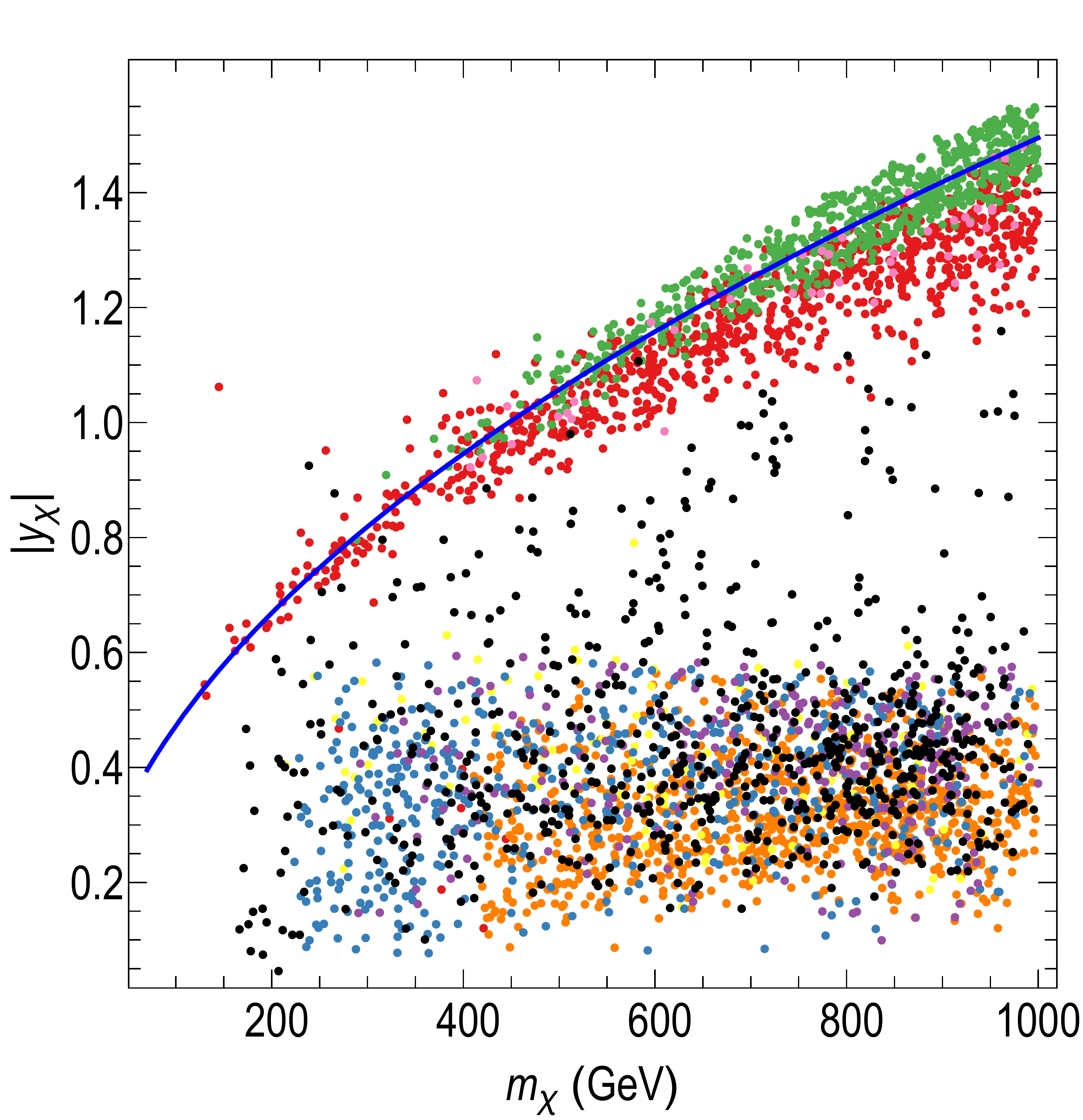}
\label{fig:scan1a}
\end{subfigure}
\hspace{1em}
\begin{subfigure}[t]{0.43\textwidth}
\includegraphics[width=\textwidth]{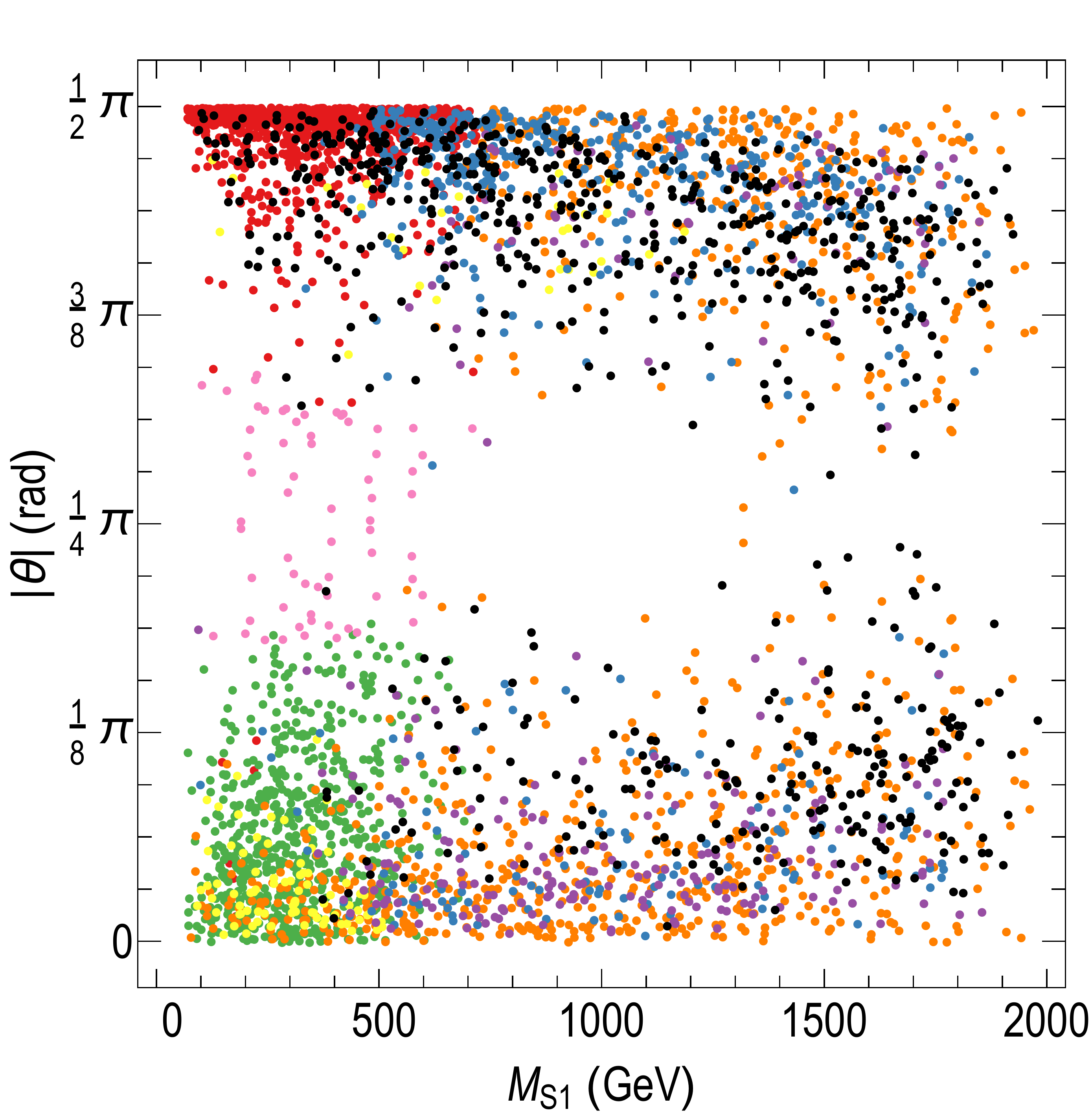}
\label{fig:scan1b}
\end{subfigure}
\vspace{0.5ex}
\begin{subfigure}[t]{0.43\textwidth}
\includegraphics[width=\textwidth]{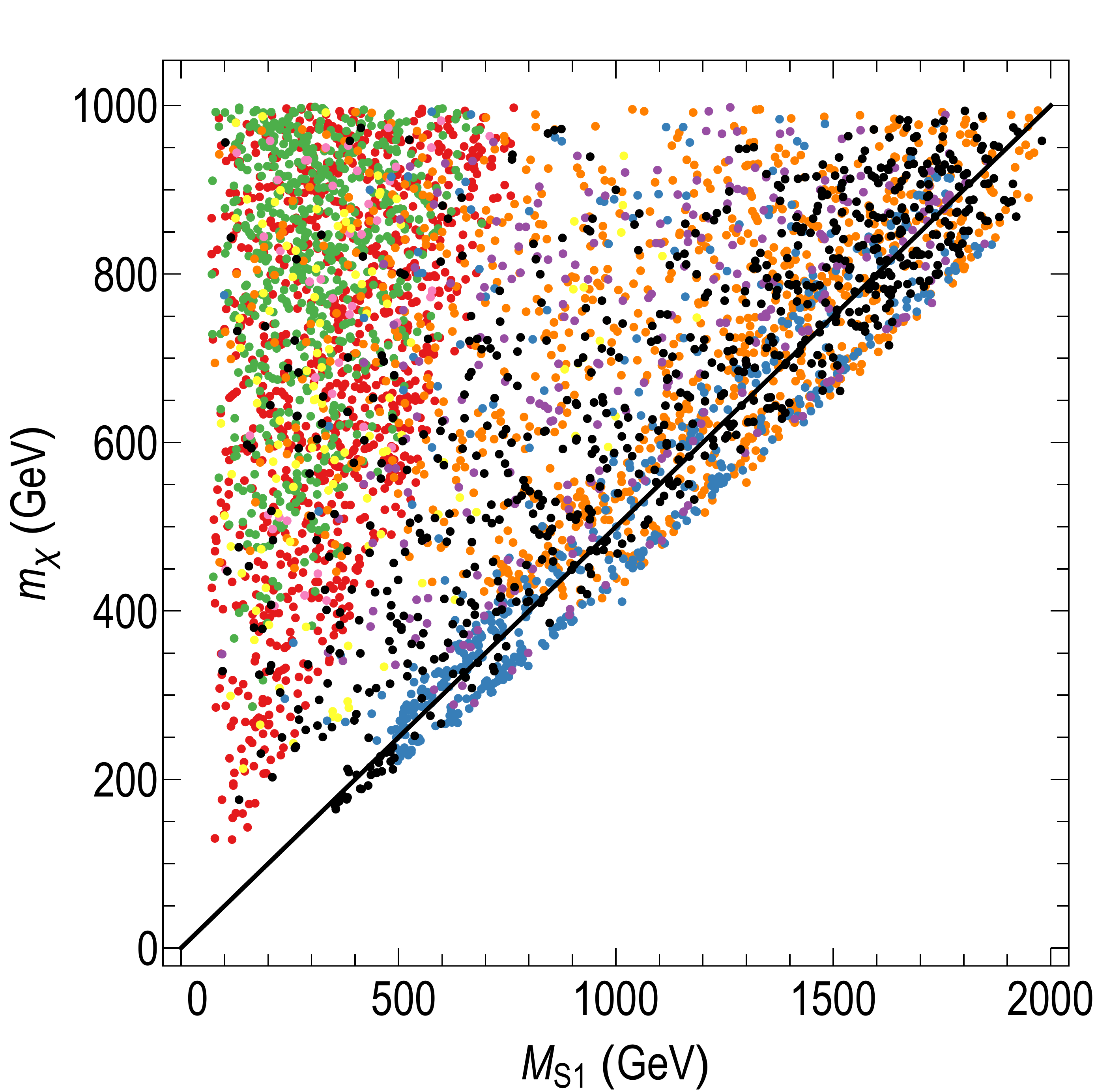}
\label{fig:scan1c}
\end{subfigure}
\hspace{1em}
\begin{subfigure}[t]{0.43\textwidth}
\includegraphics[width=\textwidth]{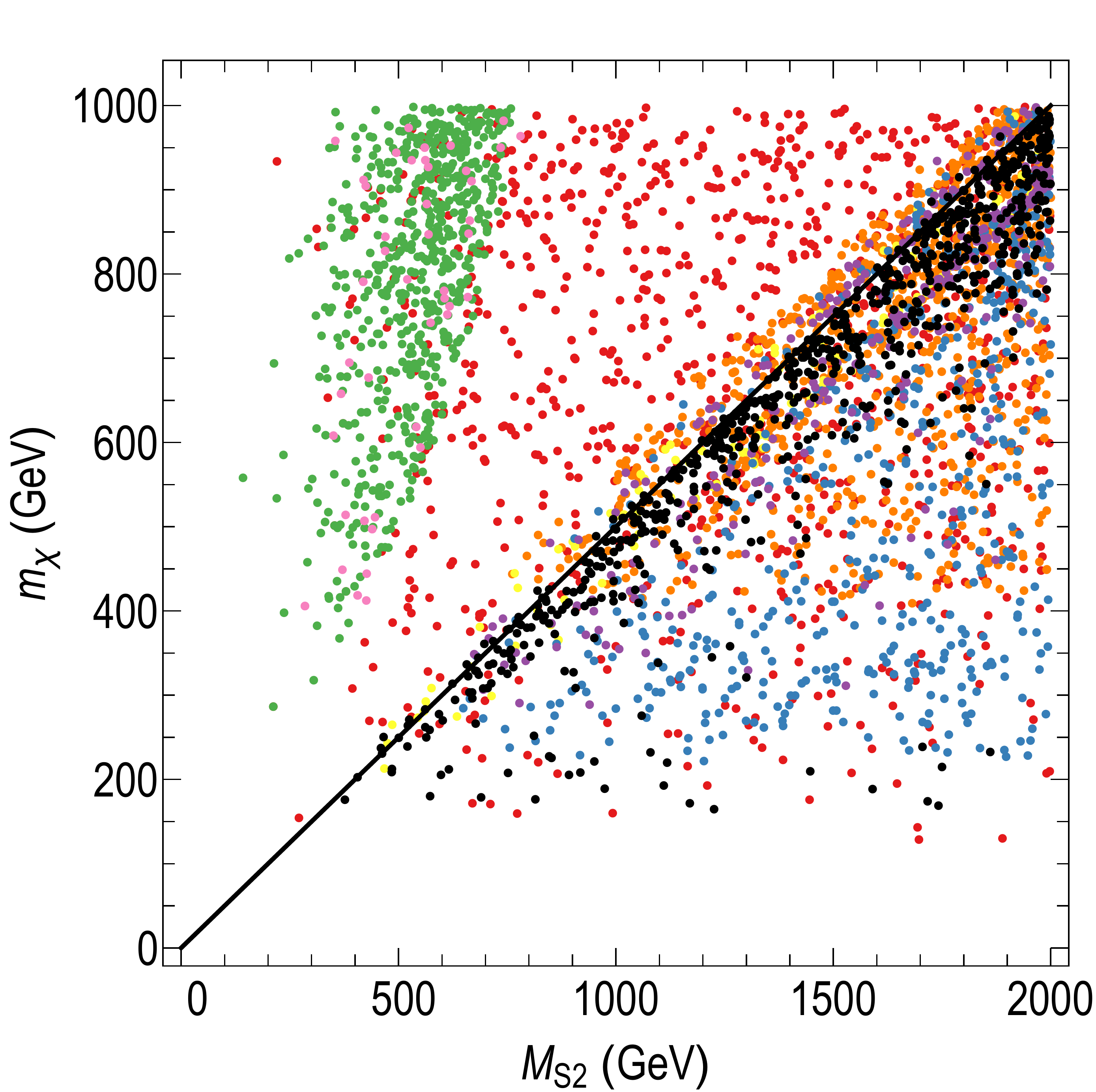}
\label{fig:scan1d}
\end{subfigure}\\
\begin{subfigure}[t]{0.86\textwidth}
\includegraphics[width=\textwidth]{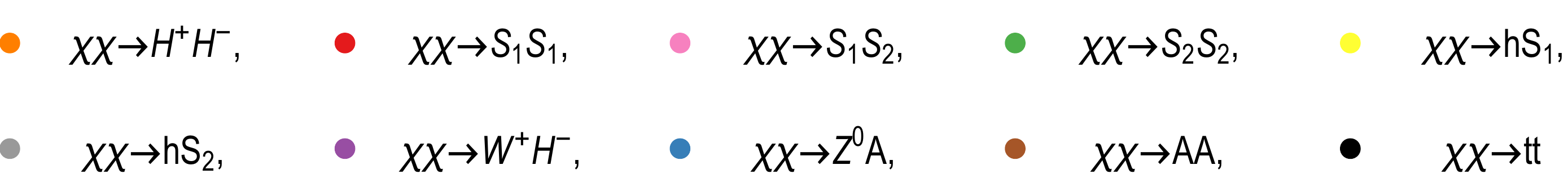}
\end{subfigure}
\caption{Points of our scan of parameter space that produce the correct relic density. The colours represent the dominant annihilation channel, as shown above. 15000 points are taken that survive the constraints, and of them only 25\% of the $H^+H^-$ channel and 10\% of the $S_1S_1$ channel are shown for clarity. The black line in the lower panels indicate $m_\chi = 2 M_{S_1, S_2}$, which is the resonance condition for the s-channel annihilation processes. The blue line in the top left panel represents the scaling expected for a cross section of $\langle \sigma v \rangle \sim y^4_\chi v^2 / 16 \pi^4 m_\chi^2$ (Eq.~\eqref{eq:sigmav}) which, in this model, applies to a pure $\overline{\chi}\chi \to S_iS_i$ scenario.}
\label{fig:scan1}
\end{figure}

\begin{figure}[ht]
\centering
\begin{subfigure}[t]{0.45\textwidth}
\includegraphics[width=\textwidth]{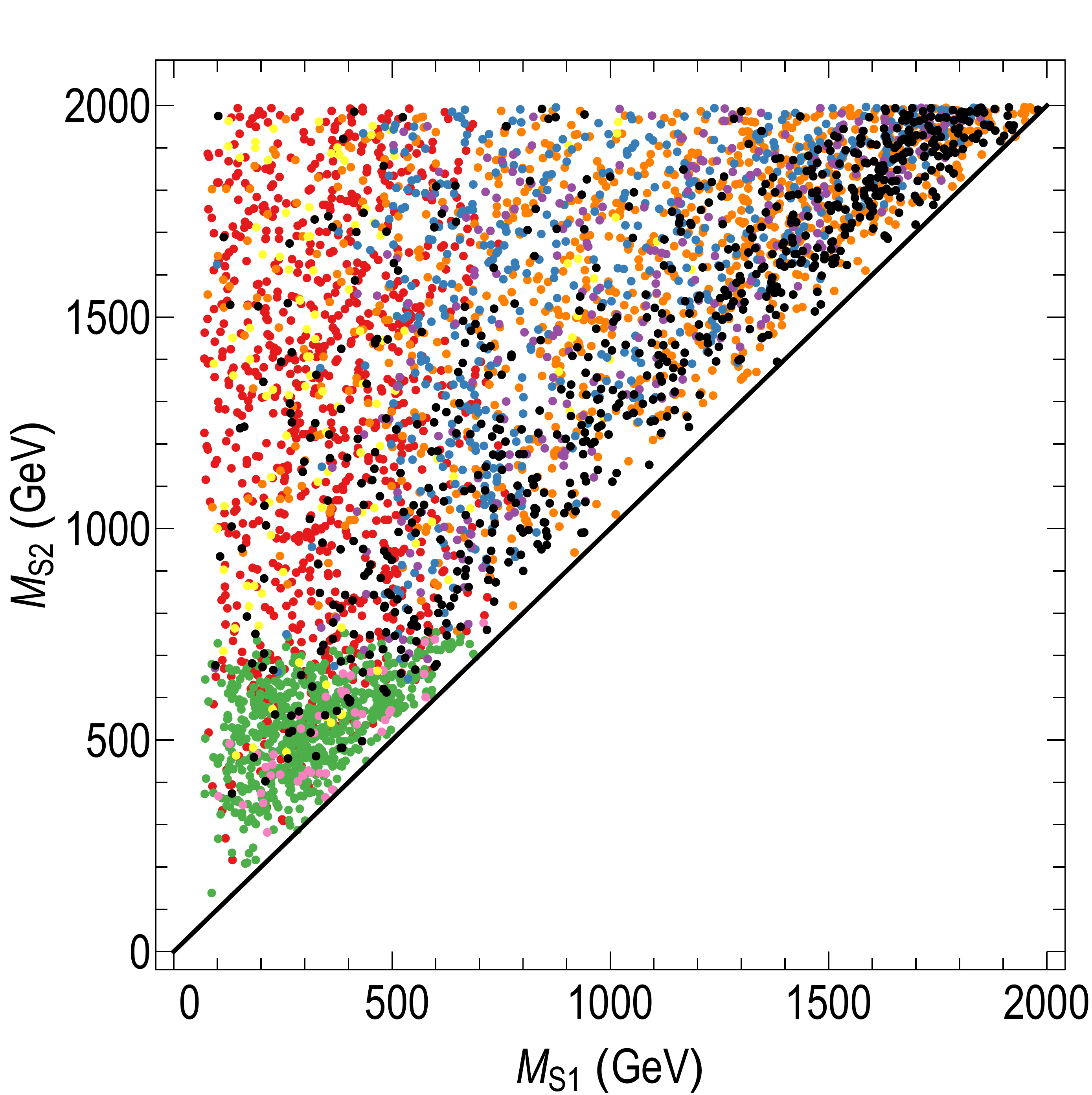}
\label{fig:scan1e}
\end{subfigure}
\hspace{1em}
\begin{subfigure}[t]{0.45\textwidth}
\includegraphics[width=\textwidth]{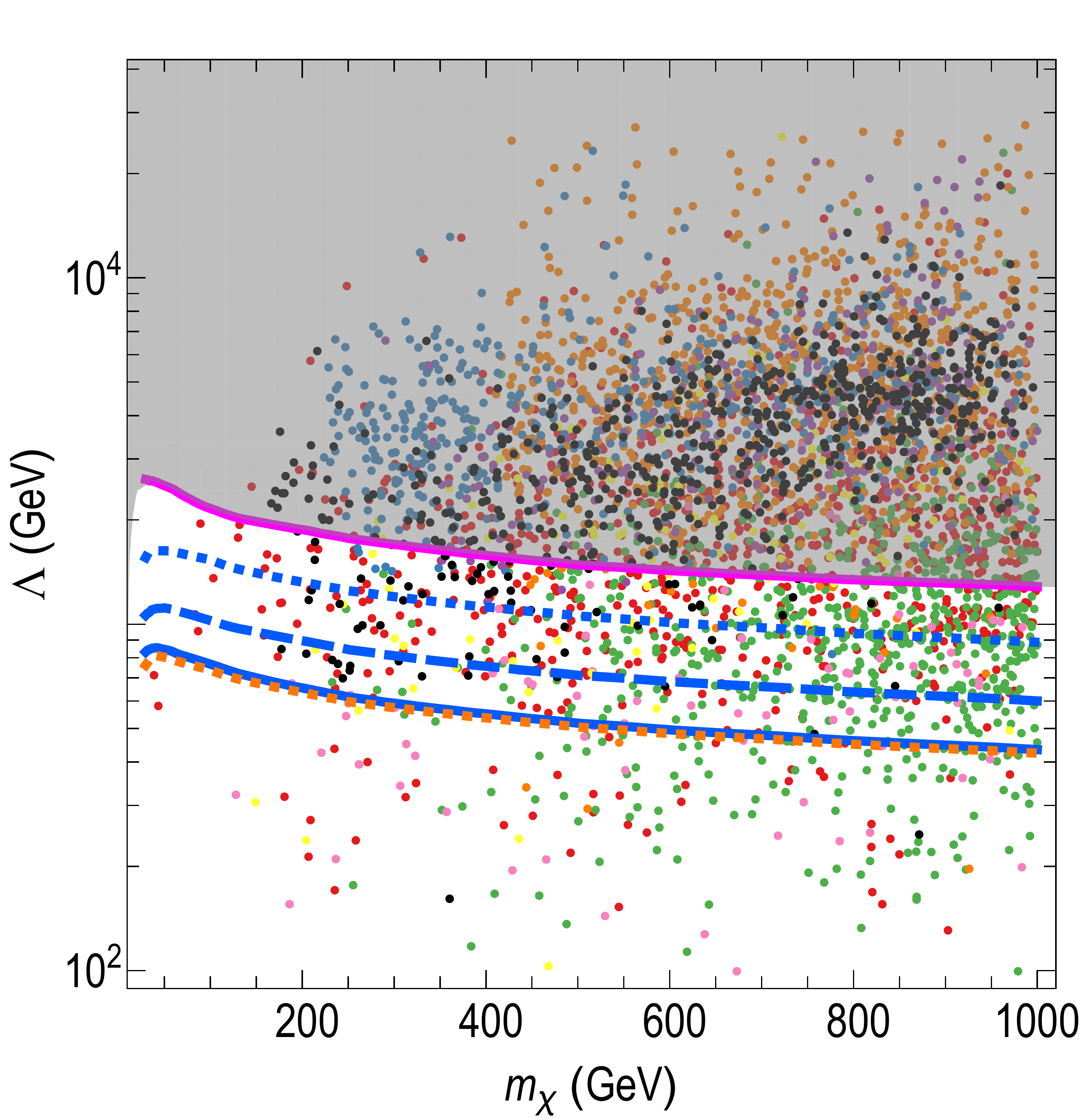}
\label{fig:scan1f}
\end{subfigure}
\begin{subfigure}[t]{0.9\textwidth}
\includegraphics[width=\textwidth]{plots/LegendH2.pdf}
\end{subfigure}
\caption{Points of our scan of parameter space that produce the correct relic density. The colours represent the dominant annihilation channel, as shown above. 15000 points are taken that survive the constraints, and of them only 25\% of the $H^+H^-$ channel and 10\% of the $S_1S_1$ channel are shown for clarity. $\Lambda$ is the effective cut-off scale for the DD effective operator; the dotted orange line represents the constraint from the LUX 2016 results \citep{Akerib:2016vxi}, the solid blue represents the XENON1T experiment \citep{Aprile:2017iyp}, the dashed blue the projection for the XENON1T experiment with $2t\cdot y$ of data taking, the dotted blue the projection for the XENONnT experiment with $20t\cdot y$ of data taking \citep{Aprile:2015uzo}, and the magenta is the DD sensitivity at the ``neutrino floor''\citep{Billard:2013qya}.}
\label{fig:scan11}
\end{figure}

The relic density is insensitive to the value of $\epsilon_d$, while the DD results depend on the relationship $\epsilon_d$ and $\epsilon_u$.  
We set $\epsilon_d = \epsilon_u$ for the scans presented in Fig.~\ref{fig:scan1} and Fig.~\ref{fig:scan11} (with the exception of the right panel of Fig.~\ref{fig:scan11}, which enforces no DD constraint and hence has no $\epsilon_d$ dependence). In Fig.~\ref{fig:scan2} and Fig.~\ref{fig:scan3} we explore additional relations between the scaling factors.

We have chosen to define $S_1$ to be the lighter of the 2 scalars, and allow $\theta$ to range from $0$ to $\pi/2$. As one can always switch the two scalars by sending $\theta\rightarrow\pi/2-\theta$, an equivalent choice would be to take $0<\theta<\pi/4$ without requiring any mass ordering.

\subsection{Scan Results and Discussion}

We present the results of the parameter scan in Fig.~\ref{fig:scan1} and \ref{fig:scan11}, where the colour of the points indicates the dominant annihilation channel. Note that, for clarity, we have plotted only 25\% of the points where the dominant annihilation channel is to $H^+H^+$, and 10\% of the points for $S_1S_1$; annihilation to these final states are by far the most frequent.

In the top left panel of Fig.~\ref{fig:scan1} we show the points of our scan that yield the correct relic density in the $m_\chi$-$y_\chi$ plane. The blue line represents the relationship between $y_\chi$ and $m_\chi$ that one would expect for a cross section of $\langle \sigma v \rangle \sim y^4_\chi v^2 / 16 \pi^4
m_\chi^2$, which applies to a pure $\overline{\chi}\chi \to S_iS_i$ scenario.
The points for which $\chi \chi \to S_1 S_1$ and $\chi \chi \to S_2 S_2$ are the dominant annihilation channels are originally centered along this line, however once we include Sommerfeld enhancement the cross sections becomes larger and thus require a lower coupling to accommodate the correct relic abundance. This results in the points where $\chi \chi \to S_1 S_1$ dominates being pushed slightly below the line\footnote{The points where $\chi \chi \to S_2 S_2$ dominates are not significantly affected, as the Sommerfeld enhancement primarily impacts the lighter of the neutral scalars, $S_1$.}. There is also a region with substantially lower values of $y_\chi$, where DM annihilations are on resonance, and in this region the dominant channels are the ones with final state gauge bosons, charged Higgs bosons and pseudoscalars. 
Finally, there is a region where the dominant annihilations are to $t\bar{t}$ pairs, with intermediate values of $y_\chi$, that appears to bridge the gap between the $S_i S_j$ region above and the resonance region below.

In the top right panel of Fig.~\ref{fig:scan1} the same points are presented in the $m_\chi$-$\theta$ plane. The red points accumulated at $\theta\sim\pi/2$ (meaning that the lighter of the two scalars is mostly the singlet) indicate a nearly secluded DM sector where the DM relic density is achieved by annihilation to the (nearly unmixed) scalar singlet, which later decays to SM particles. Final states $ZA$ and $t\bar{t}$ also seem to prefer angles in the upper half of the plot, while $S_1 h$, $H^+ H^-$, and $H^+ W^-$ seem to prefer the lower half. Finally, annihilations to $S_1 S_2$ are favored in the large mixing angle region near $\theta\sim\pi/4$; this region is poorly populated as some degree of fine tuning is necessary to achieve such mixing angles.

In the bottom left panel of Fig.~\ref{fig:scan1} we show the points in the $M_{S_1}$-$m_\chi$ plane, where the black solid line represents the s-channel resonance condition for the lighter scalar, $M_{S_1}=2m_\chi$. From this plot we can indeed verify that the $\chi \chi \to Z A$ points featuring low values of $y_\chi$ in the top left panel of Fig.~\ref{fig:scan1} are usually centered around the resonance region, while annihilations to $S_1, S_2$ and $S_1 h$ usually lie well above the $S_1$ resonance line. By comparing this plot with the one relevant for the (single) scalar mediator model in \citep{Albert:2017onk}, we observe that our allowed regions approximately correspond to the regions for which \cite{Albert:2017onk} obtained underproduction of the relic abundance. This is not unexpected, as in our case the Yukawa couplings are not fixed to one but allowed to vary, and are further suppressed by mixing angles factors.

In the bottom right panel of Fig.~\ref{fig:scan1} we instead plot in the $M_{S_2}$-$m_\chi$ plane, where the black solid line represents the s-channel resonance condition for the heavier scalar, $M_{S_2}=2m_\chi$. In this figure, we can note that points with annihilations to neutral scalars are usually above the resonance line, with the $S_2 S_2$ final states featuring the lowest values of $M_{S_2}$, and then $S_1 S_2$ and $S_1 S_1$ featuring larger and larger values. In the region where both $S_1 S_1$ and $S_2 S_2$ final states are kinematically allowed, the branching ratio to the latter is usually enhanced by larger cubic scalar couplings. 
Along the resonance line we find instead most of the points for which annihilations to $t\bar{t}$, $W^+ H^-$, and $S_1 h$ dominate (and a large portion of the $H^+ H^-$ points). Below the resonance line we find all the points where the $ZA$ channel dominates (which, as we saw in the bottom left panel of Fig.~\ref{fig:scan1}, usually have $2m_\chi \sim M_{S_1}<M_{S_2}$ and hence proceed via an $S_1$ resonance) and some of the points of the $S_1 S_1$ channel.

In the left panel of Fig.~\ref{fig:scan11} we consider the $M_{S_1}$-$M_{S_2}$ plane, where we can see that points with annihilations to scalars are all concentrated in the region $M_{S_1} \lesssim 500\GeV$, with the annihilations to $S_2 S_2$ in the $M_{S_2} \lesssim 700\GeV$ region. The points of the other channels are instead usually in the region $M_{S_1}\gtrsim 500\GeV$.

In the right panel of Fig.~\ref{fig:scan11} we show the points from the scan in the plane $m_\chi$-$\Lambda$, where $\Lambda$ is effective scale of the nucleon operator $\frac{m_N}{\Lambda^3}\bar{N}N\bar{\chi}\chi$ relevant for DD. In this plot we also show the points that are excluded by LUX \citep{Akerib:2016vxi} and XENON1T \citep{Aprile:2017iyp}. The dotted orange line indicates the current bound from LUX, while the blue solid line indicates the one from XENON1T. The dashed and dotted blue lines indicates projected exclusions limits for XENON1T after 2~ton-years and 20~ton-years of data taking, respectively, while the magenta line indicates the sensitivity corresponding to the ``neutrino floor" where low energy solar neutrinos present a significant background for DD searches~\citep{Billard:2013qya}. While the next generation DD searches will probe a significant portion of the low-mass parameter space of this model, a sizable number of points lie beyond the neutrino floor (particularly those for which $\bar{t}t$, $H^+H^-$, $ZA$, or $W^\pm H^\mp$ are the dominant annihilation modes -- i.e. the points on resonance).

\begin{figure}
\centering
\begin{subfigure}[t]{0.4\textwidth}
\includegraphics[width=\textwidth]{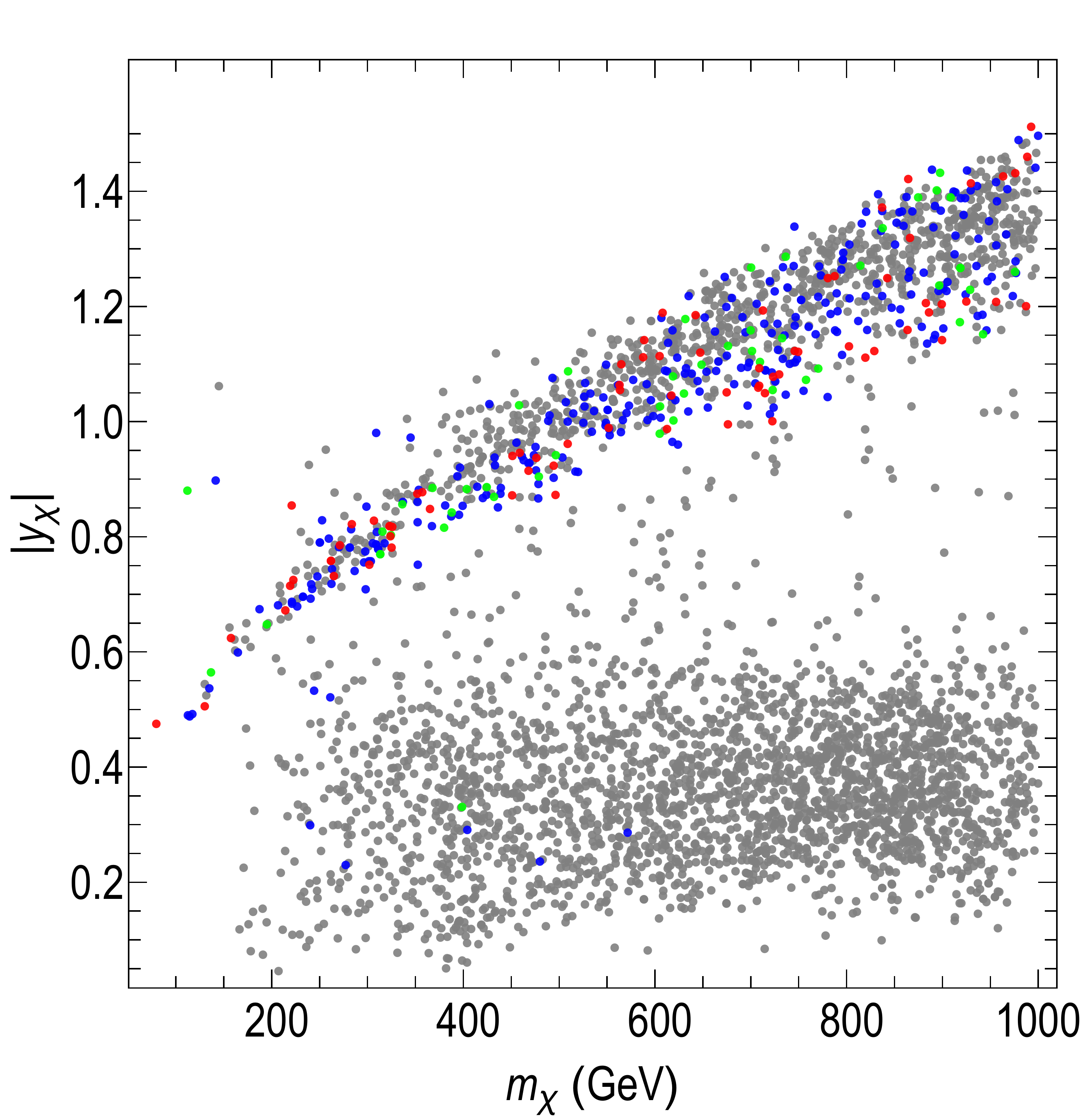}
\label{fig:scan2a}
\end{subfigure} \hspace{1em}
\begin{subfigure}[t]{.4\textwidth}
\includegraphics[width=\textwidth]{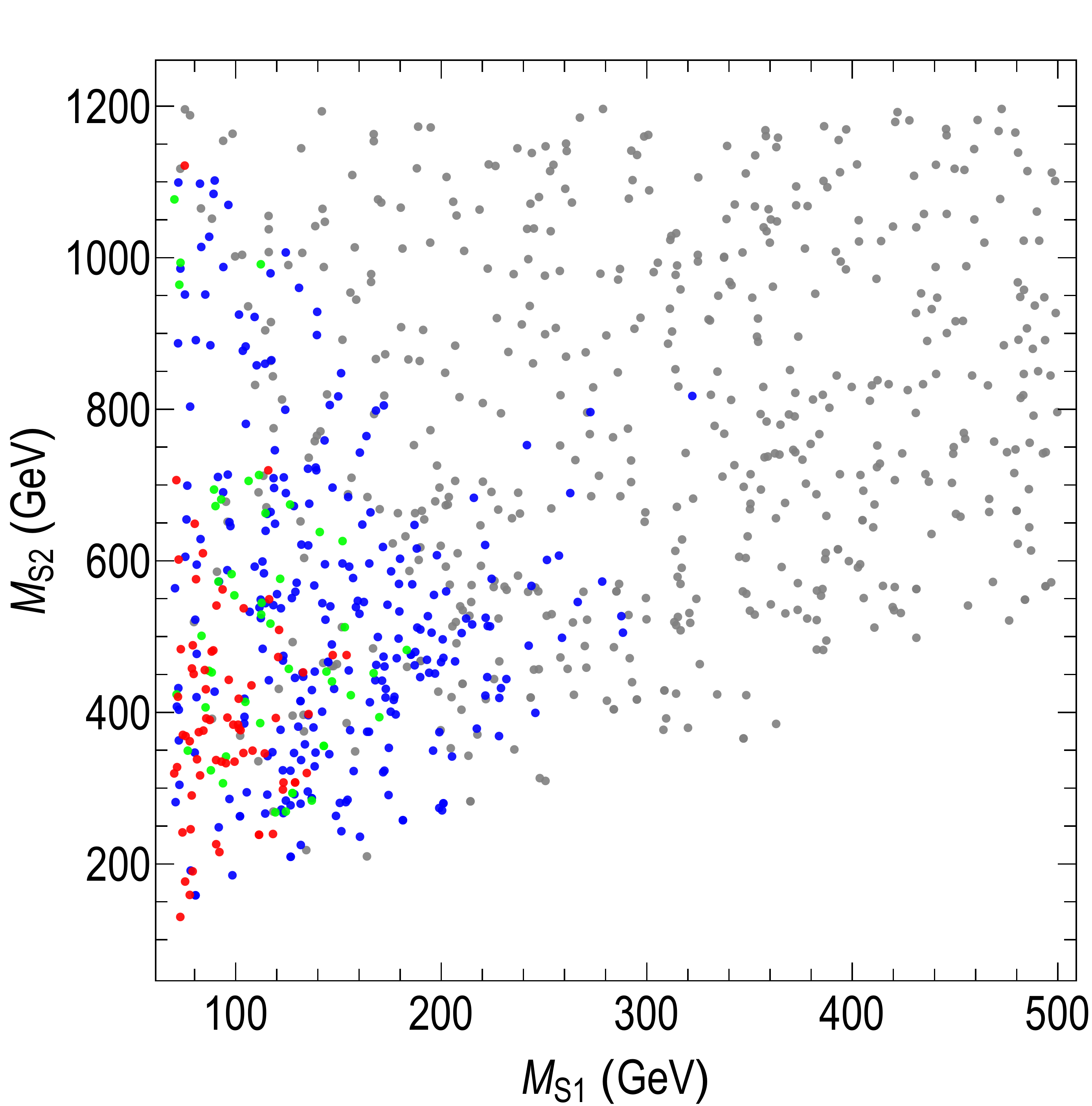}
\label{fig:scan2b}
\end{subfigure} \hspace{3em}
\\
\begin{subfigure}[t]{0.45\textwidth}
\includegraphics[width=\textwidth]{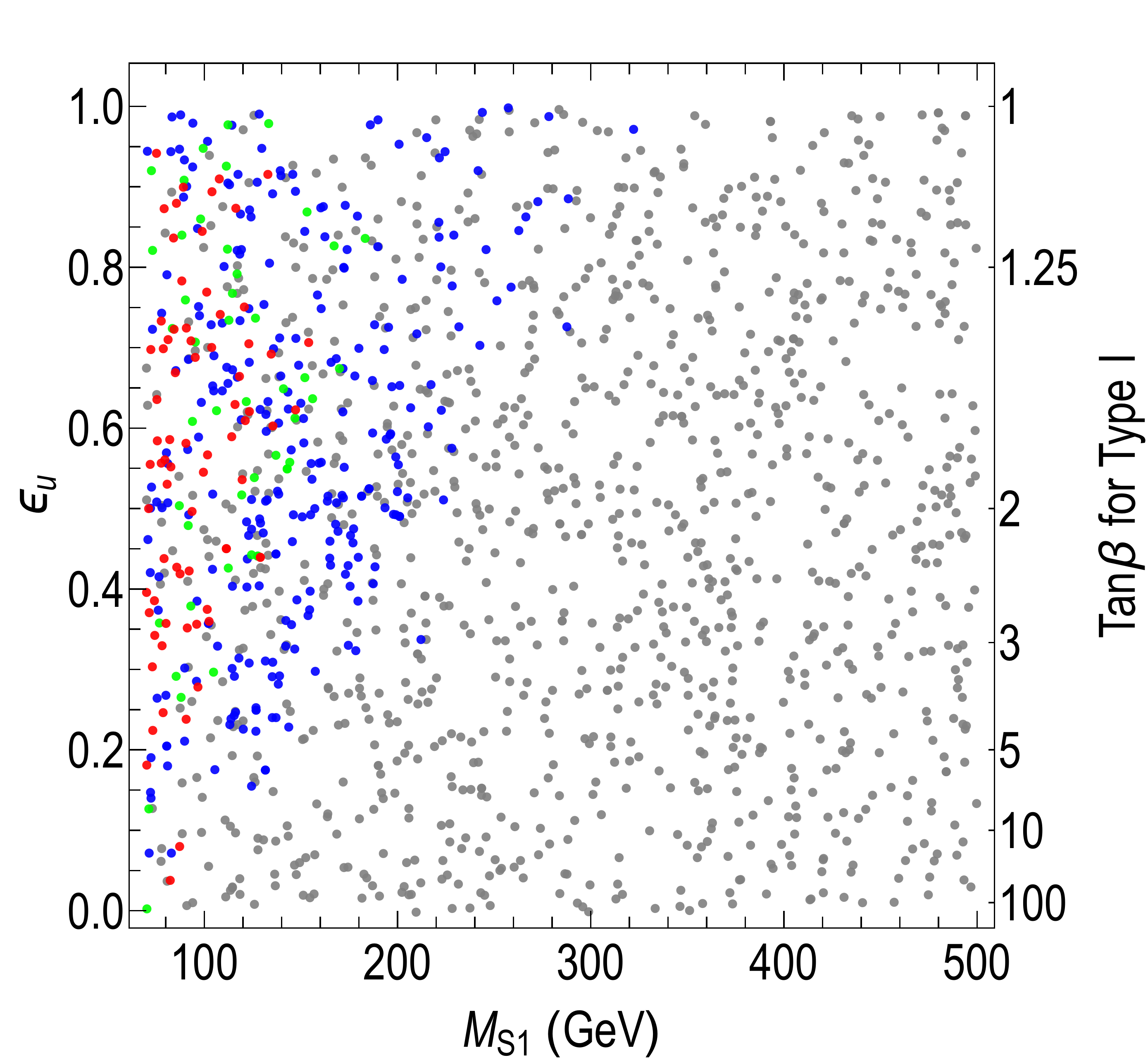}
\label{fig:scan2c}
\end{subfigure}
\caption{Direct detection constraints shown for different choices of the Yukawa structure.
Grey points are allowed if $\epsilon_d = \epsilon_u$, $\epsilon_d=0$ or $\epsilon_d=-\epsilon_u$, blue points are allowed if $\epsilon_d=0$ or $\epsilon_d=-\epsilon_u$ but excluded if $\epsilon_d=\epsilon_u$, green points are allowed if $\epsilon_d=-\epsilon_u$ but excluded in the other cases, while the red points are excluded for all cases considered.  (The case of $\epsilon_u=\epsilon_d$ corresponds to a Type I 2HDM, for which the grey points are allowed while all other points are excluded; the value of $\tan\beta\equiv1/\epsilon_{u,d}$ is shown on the RH axis of the lower plot.) The constraints were generated using the 2016~LUX \citep{Akerib:2016vxi} and XENON1T~\citep{Aprile:2017iyp} data, and tools from \citep{DelNobile:2013sia}. }
\label{fig:scan2}
\end{figure}

\begin{figure}
\centering
\begin{subfigure}[t]{0.45\textwidth}
\includegraphics[width=\textwidth]{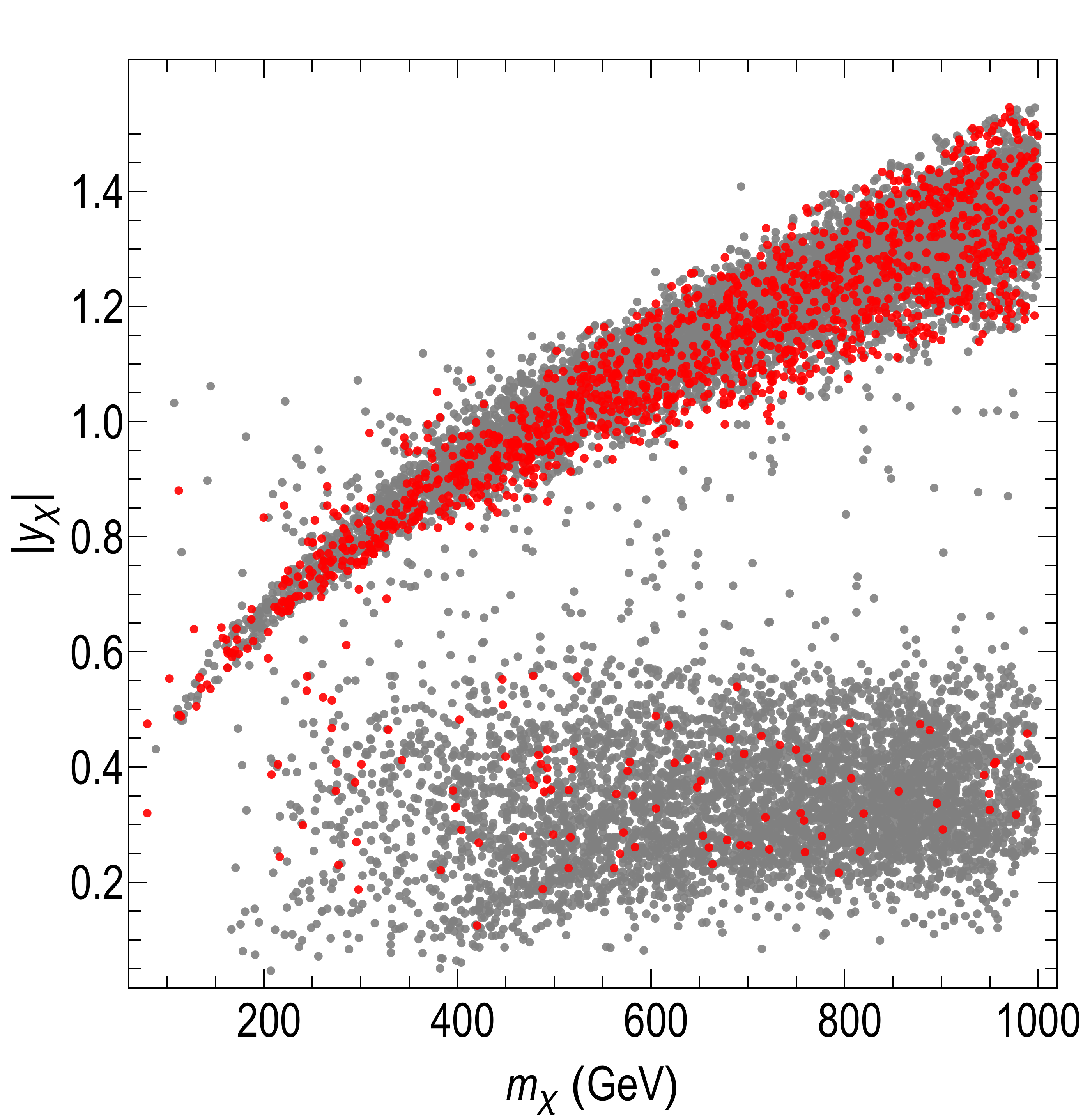}
\label{fig:scan3a}
\end{subfigure}
\begin{subfigure}[t]{.45\textwidth}
\includegraphics[width=\textwidth]{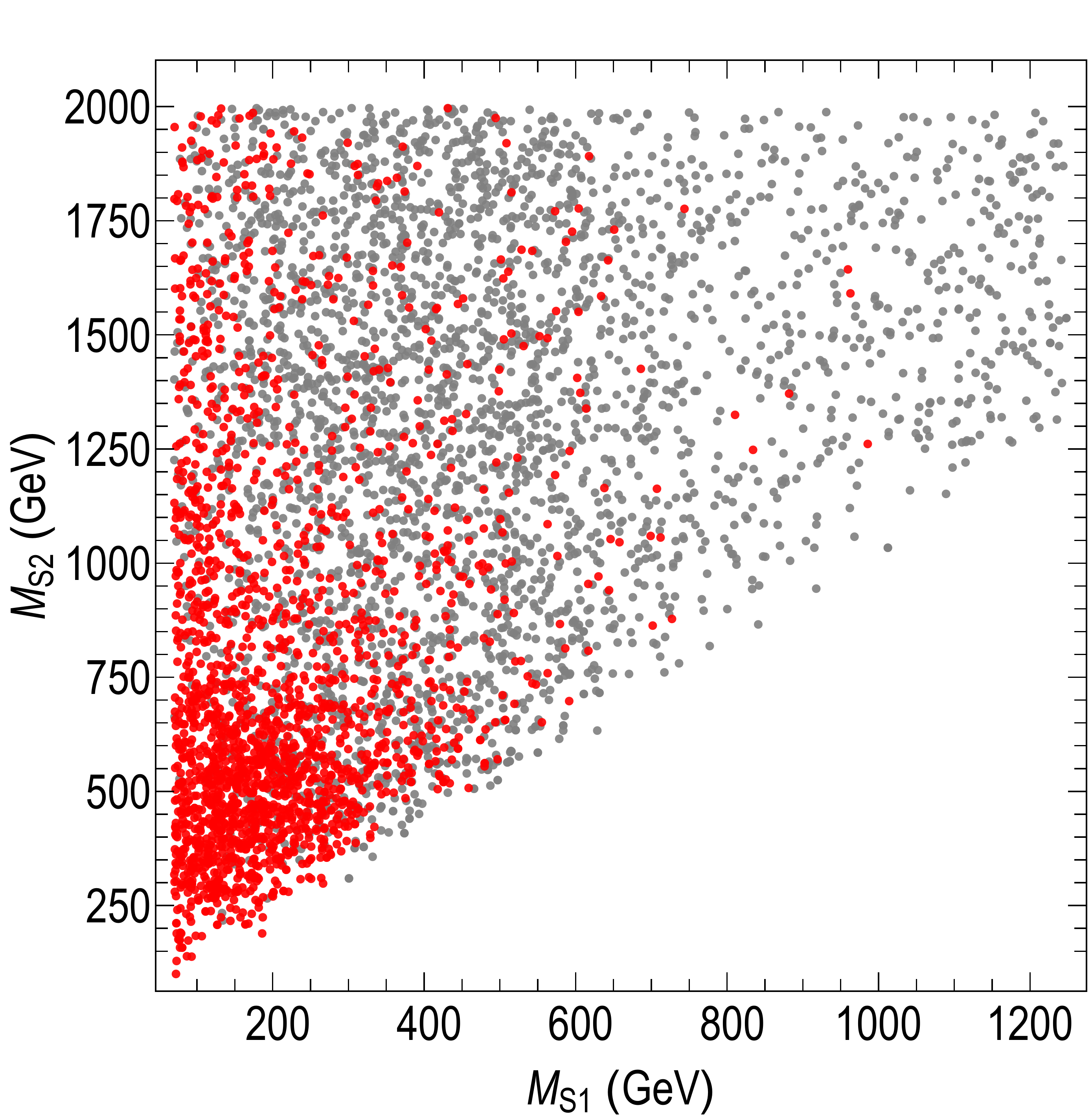}
\label{fig:scan3b}
\end{subfigure}
\\
\begin{subfigure}[t]{0.45\textwidth}
\includegraphics[width=\textwidth]{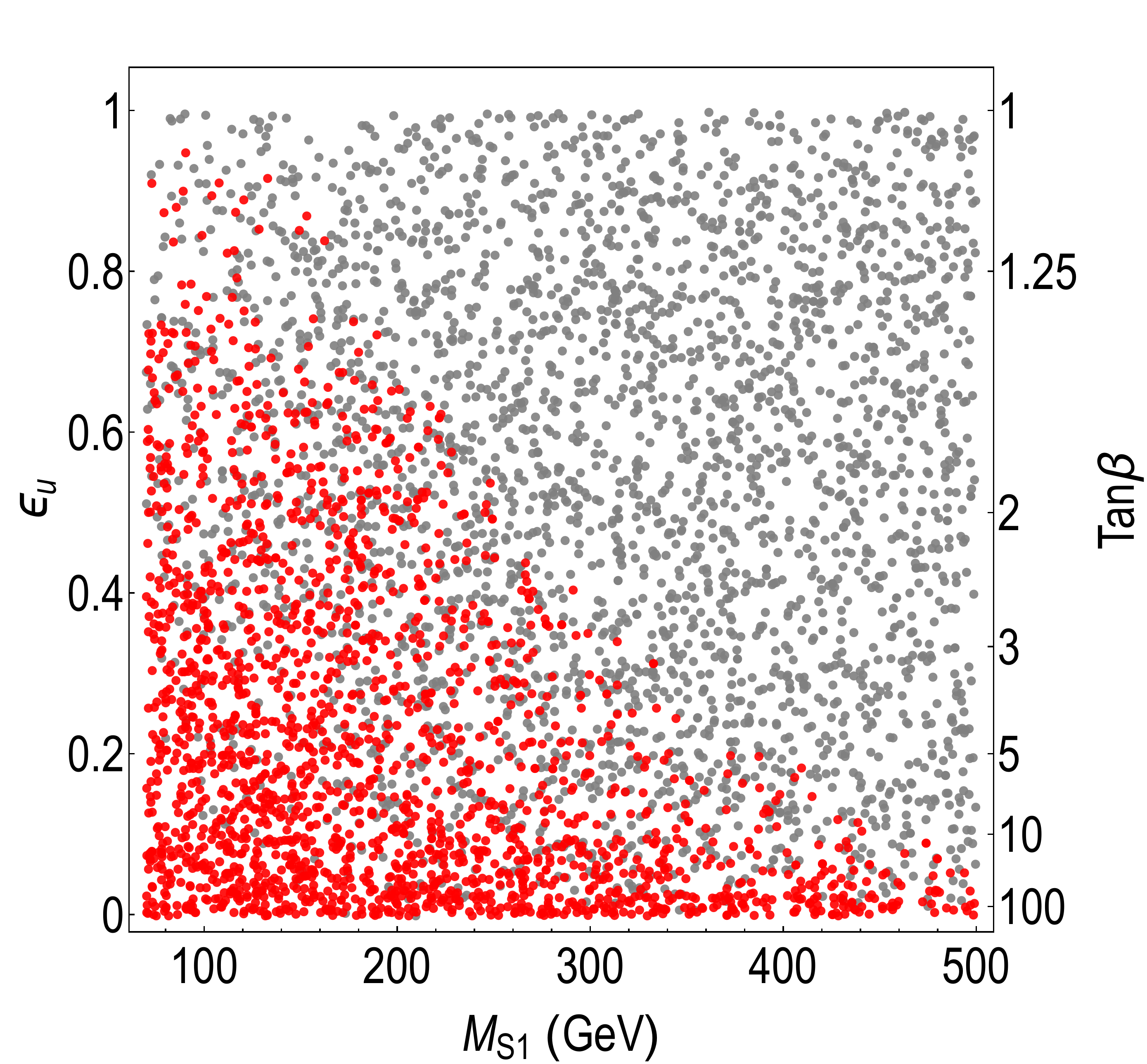}
\label{fig:scan3c}
\end{subfigure}
\caption{Direct detection constraints for a Type II 2HDM Yukawa structure. The grey points are allowed while the red points are excluded. The constraints were generated using the 2016 LUX~\citep{Akerib:2016vxi} and XENON1T~\citep{Aprile:2017iyp} data, and tools from \citep{DelNobile:2013sia}.}
\label{fig:scan3}
\end{figure}

In Fig.~\ref{fig:scan2} and  Fig.~\ref{fig:scan3} we explore the impact of different choices of Yukawa structure on the DD constraints. In  Fig.~\ref{fig:scan2} we examine which points are excluded when $\epsilon_d=\epsilon_u$, $\epsilon_d=0$, or $\epsilon_d=-\epsilon_u$ (Note that the previous figures assumed a Type I Yukawa structure, where $\epsilon_u=\epsilon_d$).
Of these choices, the condition $\epsilon_d=\epsilon_u$ leads to the largest number of excluded points (the allowed points in this case are in grey). In comparison, the condition $\epsilon_d=-\epsilon_u$ results in the weakest exclusion (grey, blue and green points allowed, with only red eliminated) due to significant $u$-$d$ interference.

In the top left panel of Fig.~\ref{fig:scan2} one can see that points which satisfy the relic density requirement via annihilation on resonance (the band below $y_\chi\simeq 0.5$) are are currently out of reach for DD experiments, which is because they typically feature values of the couplings that are too small.  In the top right panel of Fig.~\ref{fig:scan2} we can see that DD is most sensitive to points with low $M_{S_1}$, and an $M_{S_2}$ value that is not too large. The latter condition comes from the fact that the larger the mass difference $M_{S_2}^2-M_{S_1}^2$, the smaller the mixing angle, resulting in an effective suppression of the quark couplings. We see that DD is mostly sensitive the low values of $M_{S_1}\lesssim 300\GeV$, as expected. These points are usually characterized by a mixing angle close to $\theta=\pi/2$ (see upper right panel of Fig.~\ref{fig:scan1}) and lie in the region where Sommerfeld enhancement is $\mathcal{O}(1)$ but not negligible. Finally, from the bottom panel of Fig.~\ref{fig:scan2} we can see that, even though DD is more sensitive to larger values of $\epsilon_u$, this is indeed not the dominant parameter (rather, $M_{S_1}$ is). One is able to exclude some of the points down to values as low as $\epsilon_u\sim 0.2$ (i.e. $\tan\beta\sim3-5$ in a Type I model).

Finally, in Fig.~\ref{fig:scan3} we show the same plots for a Type II (or Y) 2HDM. In these plots, grey points are allowed while red points are excluded. In the case of Type II, we see that many more points are excluded. This happens because large values of $\tan\beta$ (corresponding to low values of $\epsilon_u$) enhance the couplings to $d,s$ quarks, resulting in a much larger nucleon operator coefficient in Eq.~\ref{eq:DDcoeff}. Note that for the Type II 2HDM, the up and down quarks have Yukawa couplings of opposite sign, and therefore we have negative interference between flavours. Such interference is maximal near $\tan\beta\sim0.8$ where indeed the limits are very weak. 
These facts together explain why the excluded points in the Type II example of Fig.~\ref{fig:scan3} are concentrated near $\epsilon_u=0$ (large $\tan\beta$) while in the Type I example of Fig.~\ref{fig:scan2} they are more concentrated near $\epsilon_u\sim 1$.

\section{Conclusions}
\label{sec:conclusions}

We have analysed the scenario in which DM communicates with the visible sector via an s-channel scalar mediator. In this situation, a model with a single mediator is inadequate; the minimal self-consistent, gauge invariant model involves the mixing of a singlet scalar with the neutral CP-even component of a Higgs doublet. Because mixing with the SM Higgs doublet is constrained, we have instead considered the mixing of the singlet scalar with the additional scalar in a two Higgs doublet model. Even though this model features an undesirably larger parameter space than the Simplified Models previously considered, there are some important new features that arise.

Due to the singlet-doublet mixing, the scalar mediators automatically couple to the electroweak gauge bosons and the SM Higgs boson, as well as the charged scalar and pseudoscalar that arise from the 2HDM. This opens up a number of new DM annihilation channels. While annihilation to the scalar mediators ($S_1 S_2$, $S_1 S_2$ and $S_2 S_2$) typically dominates when kinematically allowed, there is a sizable portion of parameter space for which annihilation to $H^+H^-$ has the largest branching ratio, while channels such as $ZA$, $W^\pm H^\mp$, $hS_1$ and $\bar{t}t$ tend to dominate when the annihilation proceeds via an s-channel resonance. Notably, the DM relic abundance can be adequately achieved via freezeout without any significant coupling of the DM to SM fermions. This implies the strongest beyond Standard Model signals might not be in interactions with SM fermions, but with the Higgs or gauge bosons. This is a clear difference from the one-mediator simplified models.

There will always be two scalar mediators, and they will in general interfere. Interference effects have been shown to be important in DD, as they generate blind spots when the mediators have similar masses. This interference between the two mediators will be present in every process, so will also affect relic density production. There is, however, an additional source of interference that affects DD alone: the interference between different flavours. This happens if up and down quarks have different-sign Yukawa couplings as, for example, in Type II 2HDMs. As the DM annihilation cross section has a negligible dependence on the down quark Yukawas, any parameters that produce the correct relic abundance can avoid DD constraints with a suitable choice of down quark Yukawa couplings. In this regard, the Yukawa freedom introduced by the 2HDM is an important feature of the model.

We performed a scan over model parameters, to determine values which produce the correct relic abundance while satisfying DD, flavour, precision electroweak and other constraints. We found that the model can naturally produce the correct relic density with couplings in the range $\lambda_i \in [ 0.1,1 ] $, for DM masses $m_\chi \gtrsim 100\GeV$. This can be accomplished independent of the Yukawa structure chosen, as the dominant annihilation channels are typically those to scalars and hence the most important parameters are the DM mass $m_\chi$ and Yukawa $y_\chi$, and scalar masses $M_{S_1,S_2}$.
We also examined the DD constraints on parameters with viable relic density production. Here we found that, in the optimistic scenario where one has constructive interference between flavours ($\epsilon_u=\epsilon_d=\cot\beta$ for Type I) DD is able to probe only the parameters where one of the scalars is below 250~GeV. Alternatively, for Type II Yukawa structures, the DD limits are very weak around $\tan\beta \sim 1$ (where $\epsilon_u = - \epsilon_d$) but much stronger at large values of $\tan\beta = 1/\epsilon_u$ where the enhanced down quark couplings dominate the scattering cross section.  In this case DD has sensitivity to scalar masses as large as $500$~GeV.

\section*{Acknowledgements}
NFB and GB were supported in part by the Australian Research Council, and 
and IWS by the Commonwealth of Australia. 
Feynman diagrams were drawn using TikZ-Feynman \citep{Ellis:2016jkw}.

\newpage
\appendix

\section{Relic Density Calculation}
\label{sec:relicdensity}

\subsection{Relic Density Cross Sections}
\label{sec:relicdensitycs}

The expressions for the DM annihilation cross sections $\bar{\chi} \chi \to $ final states are shown below. Note that the width of the mediators have been omitted from these expressions, but can be easily restored by substitutions such as $\left(M_{S_i}^2-s\right)^2 \rightarrow \left(\left(M_{S_i}^2-s\right)^2+M_{S_i}^2\Gamma_i^2\right)$. 
\bea
\sigma_{H^+ H^-} &=& \frac{\hat{\lambda}_{HHs}^2 v_s^2 y_\chi^2 \sqrt{1-\frac{4
   M_{H^+}^2}{s}} \sqrt{1-\frac{4
   m_\chi^2}{s}} \left(\cos (2 \theta )
   \left(M_{S_1}^2-M_{S_2}^2\right)+M_{S_1}^2+M_{S_2}^2-2 s\right)^2}{128 \pi
    \left(M_{S_1}^2-s\right)^2 \left(M_{S_2}^2-s\right)^2}, \quad \\
\sigma_{AA} &=& \frac{\hat{\lambda}_{HHS}^2 v_s^2 y_\chi^2 \sqrt{1-\frac{4
   M_{A}^2}{s}} \sqrt{1-\frac{4
   m_\chi^2}{s}} \left(\cos (2 \theta )
   \left(M_{S_1}^2-M_{S_2}^2\right)+M_{S_1}^2+M_{S_2}^2-2 s\right)^2}{256 \pi
    \left(M_{S_1}^2-s\right)^2 \left(M_{S_2}^2-s\right)^2},\\
\sigma_{S_i S_j} &=& \frac{y_\chi^2\sqrt{1-\frac{\left(M_{S_i}-M_{S_j}\right)^2}{s}}\sqrt{1-\frac{\left(M_{S_i}+M_{S_j}\right)^2}{s}}\sqrt{1-\frac{4m_\chi^2}{s}}}{64\pi}\left(1-\frac{\delta_{ij}}{2}\right)\nn\\
&&\times \int_{-1}^1 d\cos\theta_h \left(y_{ij}^2+2y_{ij} z_{ij} +z_{ij}^2+k\tan^2 \theta_h\hat{z}_{ij}^2\right) \nn\\
&=&\frac{y_\chi^2\sqrt{1-\frac{\left(M_{S_i}-M_{S_j}\right)^2}{s}}\sqrt{1-\frac{\left(M_{S_i}+M_{S_j}\right)^2}{s}}\sqrt{1-\frac{4m_\chi^2}{s}}}{64\pi}\left(1-\frac{\delta_{ij}}{2}\right)\nn\\
&&\times\int_{-1}^1 d\cos\theta_h \left(\frac{\left(t_{ij}^2y_{ij}+\cos^2\theta_h(x_{ij}-y_{ij})-\hat{x}_{ij}\right)^2+k\cos^2\theta_h\sin^2\theta_h x_{ij}^2}{\left(t_{ij}^2-\cos^2\theta_h\right)^2}\right), \\
\sigma_{H^+ W^-} &=& 
 \frac{g^2 y_\chi^2 \sin^2 (2\theta)}{512 \pi } 
 \frac{\left(M_{S_1}^2-M_{S_2}^2\right)^2 \sqrt{1-\frac{4 m_\chi^2}{s}} \left(M_{H^+}^4 \!\!-2 M_{H^+}^2 \!\! \left(M_{W}^2 +s\right)+\left(M_{W}^2 -s\right)^2\right)^{3/2}}{M_{W}^2s \left(M_{S_1}^2-s\right)^2 \left(M_{S_2}^2-s\right)^2}, \quad \quad \\
\sigma_{A Z} &=& \frac{g^2 y_\chi^2 \sin^2 (2\theta)}{512 \pi }  \frac{\left(M_{S_1}^2-M_{S_2}^2\right)^2 \sqrt{1-\frac{4 m_\chi^2}{s}} \left(M_A^4-2 M_A^2 \left(M_{Z}^2+s\right)+\left(M_{Z}^2-s\right)^2\right)^{3/2}}{M_{W}^2 s \left(M_{S_1}^2-s\right)^2 \left(M_{S_2}^2-s\right)^2},\\
\sigma_{S_i h} &=& \frac{y_\chi^2\sqrt{1-\frac{\left(M_{S_i}-M_h\right)^2}{s}}\sqrt{1-\frac{\left(M_{S_i}+M_h\right)^2}{s}}\sqrt{1-\frac{4m_\chi^2}{s}}}{32\pi} y_{ih}^2,\\
\sigma_{t\bar{t}} &=& \frac{N_c y_\chi^2 y_t^2 \epsilon_u^2\sin^2\left(2\theta\right)}{128\pi}\sqrt{1-\frac{4m_t^2}{s}}\sqrt{1-\frac{4m_\chi^2}{s}}\frac{\left(s-4m_t^2\right)\left(M_{S_1}^2-M_{S_2}^2\right)^2}{\left(s-M_{S_1}^2\right)^2\left(s-M_{S_2}^2\right)^2},
\eea
where
\bea
f_1 &=& \sin\theta, \quad f_2 = \cos\theta,\\
x_{ij} &=& \frac{4y_\chi f_i f_j m_\chi s}{\left(s-4m_\chi^2\right)\left(s-(M_{S_i}-M_{S_j})^2\right)},\\
\hat{x}_{ij} &=& \frac{8y_\chi f_i f_j m_\chi(s-M_{S_i}^2-M_{S_j}^2)s}{\left(s-(M_{S_i}+M_{S_j})^2\right)\left(s-(M_{S_i}-M_{S_j})^2\right)\left(s-4m_\chi^2\right)},\\
y_{ij} &=& -\left(\frac{\mu_{ij1}\sin\theta}{s-M_{S_1}^2}+\frac{\mu_{ij2}\cos\theta}{s-M_{S_2}^2}\right),\\
z_{ij} &=& \frac{x_{ij} \cos^2\theta_h-\hat{x}_{ij}}{t_{ij}^2-\cos^2\theta_h},\\
\hat{z}_{ij} &=& \frac{x_{ij} \cos^2\theta_h}{t_{ij}^2-\cos^2\theta_h},\\
t_{ij} &=& \frac{\sqrt{s}\left(s-(M_{S_i}^2+M_{S_j}^2)\right)}{\sqrt{s-(M_{S_i}-M_{S_j})^2}\sqrt{s-(M_{S_i}+M_{S_j})^2}\sqrt{s-4m_\chi^2}},\\
k &=& \frac{s}{4m_\chi^2},
\eea
and
\bea
\mu_{111} &=& -\frac{3\sin\theta}{v_s}\left(\sin^2\theta M_{S_1}^2 +\cos^2\theta \hat{\lambda}_{HHs} v_s^2  \right),\\
\mu_{112} &=& -\frac{\cos\theta}{v_s}\left(\sin^2\theta \left(2M_{S_1}^2+M_{S_2}^2\right) +\left(1-3\sin^2\theta\right) \hat{\lambda}_{HHs} v_s^2  \right),\\
\mu_{122} &=& -\frac{\sin\theta}{v_s}\left(\cos^2\theta \left(M_{S_1}^2+2M_{S_2}^2\right) +\left(1-3\cos^2\theta\right) \hat{\lambda}_{HHs} v_s^2  \right),\\
\mu_{222} &=& -\frac{3\cos\theta}{v_s}\left(\cos^2\theta M_{S_2}^2 +\sin^2\theta \hat{\lambda}_{HHs} v_s^2  \right),\\
\mu_{11h} &=& -\frac{1}{2v}\sin^2(2\theta)\left(M_{S_1}^2-M_{S_2}^2\right) -\frac{m_h^2}{v}\cos^2\theta,\\
\mu_{12h} &=& -\frac{1}{4v}\sin(4\theta)\left(M_{S_1}^2-M_{S_2}^2\right) +\frac{m_h^2}{2v}\sin(2\theta),\\
\mu_{22h} &=& \frac{1}{2v}\sin^2(2\theta)\left(M_{S_1}^2-M_{S_2}^2\right) -\frac{m_h^2}{v}\sin^2\theta.
\eea

The expressions for $\sigma v$ are related to the cross sections $\sigma$ by
\be
\sigma v_{\chi\bar{\chi}\rightarrow X} = \sigma_{\chi\bar{\chi}\rightarrow X} \frac{\sqrt{s(s-4m_\chi^2)}}{s-2m_\chi^2}.
\ee
Usually (away from resonances and thresholds~\citep{Griest:1989wd,Griest:1990kh,Gondolo:1990dk}), it is safe to expand in powers of the relative velocity by setting the center of mass energy to be
\be
s = 2m_\chi^2\left(1+\frac{1}{\sqrt{1-v^2}}\right),
\ee
and use the non-relativistic expansion
\bea
\sigma v &=& a + b v^2  + {\cal O}(v^4), \\ 
\langle \sigma v \rangle &=& a + 6b/x_F.
\eea
For fermions annihilating only to or through scalar mediators, all s-wave coefficients are zero ($a=0$). The p-wave coefficients for the $\bar{\chi}\chi \to$ final states give the following expressions for the $b$ terms:
\bea
b_{H^+ H^-} &=& \frac{\hat{\lambda}_{HHS}^2 v_s^2 y_\chi^2 \sqrt{1-\frac{M_{H^+}^2}{m_\chi^2}} \left(\cos (2 \theta )
   \left(M_{S_1}^2-M_{S_2}^2\right)+M_{S_1}^2+M_{S_2}^2-8 m_\chi^2\right)^2}{256 \pi  \left(M_{S_1}^2-4 m_\chi^2\right)^2 \left(M_{S_2}^2-4 m_\chi^2\right)^2},\\
b_{AA} &=& \frac{\hat{\lambda}_{HHS}^2 v_s^2 y_\chi^2 \sqrt{1-\frac{M_{A}^2}{m_\chi^2}} \left(\cos (2 \theta )
   \left(M_{S_1}^2-M_{S_2}^2\right)+M_{S_1}^2+M_{S_2}^2-8 m_\chi^2\right)^2}{512 \pi  \left(M_{S_1}^2-4 m_\chi^2\right)^2 \left(M_{S_2}^2-4 m_\chi^2\right)^2},\\
b_{S_1 S_1} &=& \frac{y_\chi^2\sqrt{1-\frac{M_{S_1}^2}{m_\chi^2}}}{128\pi\left(2m_\chi^2-M_{S_1}^2\right)^4} \bigg[ \frac{32}{9}y_\chi^2 \sin^4\theta m_\chi^2\left(m_\chi^2-M_{S_1}^2\right)^2 \nn \\
&+& \left(\frac{4}{3}y_\chi\sin^2\theta m_\chi\left(5m_\chi^2-2M_{S_1}^2\right)-\hat{y}_{11}\left(2m_\chi^2-M_{S_1}^2\right)^2\right)^2\bigg],\\
b_{S_2 S_2} &=& \frac{y_\chi^2\sqrt{1-\frac{M_{S_2}^2}{m_\chi^2}}}{128\pi\left(2m_\chi^2-M_{S_2}^2\right)^4} \bigg[\frac{32}{9}y_\chi^2 \cos^4\theta m_\chi^2\left(m_\chi^2-M_{S_2}^2\right)^2 \nn \\
&&+ \left(\frac{4}{3}y_\chi\cos^2\theta m_\chi\left(5m_\chi^2-2M_{S_2}^2\right)-\hat{y}_{22}\left(2m_\chi^2-M_{S_2}^2\right)^2\right)^2\bigg],\\
b_{S_1 S_2} &=& \frac{y_\chi^2\sqrt{1-\frac{(M_{S_1}-M_{S_2})^2}{4m_\chi^2}}\sqrt{1-\frac{(M_{S_1}+M_{S_2})^2}{4m_\chi^2}}}{64\pi\left(4m_\chi^2-M_{S_1}^2-M_{S_2}^2\right)^4}\nn\\
&&\times\left(\frac{128}{9}y_\chi^2 \sin^2(2\theta) m_\chi^6 \left(1-\frac{(M_{S_1}-M_{S_2})^2}{4m_\chi^2}\right)^2\left(1-\frac{(M_{S_1}+M_{S_2})^2}{4m_\chi^2}\right)^2\right.\nn\\
&& \quad +\left( \frac{8}{3}y_\chi\sin(2\theta) m_\chi^3\left(1-\frac{(M_{S_1}-M_{S_2})^2}{4m_\chi^2}\right)\left(1-\frac{(M_{S_1}+M_{S_2})^2}{4m_\chi^2}\right)\right.\nn\\
&& \quad +\left.\left(4m_\chi^2-M_{S_1}^2-M_{S_2}^2\right)\left(\left(4m_\chi^2-M_{S_1}^2-M_{S_2}^2\right)\hat{y}_{12}-4y_\chi \sin(2\theta)m_\chi\right)\bigg)^2\right),\\
b_{H^+ W^-} &=& \frac{g^2 y_\chi^2 \sin^2 (2\theta)}{4096\pi}
\frac{ \left(M_{S_1}^2-M_{S_2}^2\right)^2
   \left(M_{H^+}^4 \!\! -2 M_{H^+}^2 \!\!\left(4 m_\chi^2 \!+ \! M_{W}^2\right)+\left(M_{W}^2-4 m_\chi^2\right)^2\right)^{3/2}}{ m_\chi^2 M_{W}^2 \left(M_{S_1}^2-4 m_\chi^2\right)^2 \left(M_{S_2}^2-4
   m_\chi^2\right)^2}, \quad\quad\, \\
b_{A Z} &=& \frac{g^2 y_\chi^2 \sin^2 (2\theta)}{4096\pi} 
\frac{ \left(M_{S_1}^2-M_{S_2}^2\right)^2
   \left(M_A^4-2 M_A^2 \left(4 m_\chi^2+M_{Z}^2\right)+\left(M_{Z}^2-4 m_\chi^2\right)^2\right)^{3/2}}{ m_\chi^2 M_{W}^2 \left(M_{S_1}^2-4 m_\chi^2\right)^2 \left(M_{S_2}^2-4
   m_\chi^2\right)^2},\\
b_{S_i h} &=& \frac{y_\chi^2}{62\pi}\sqrt{1-\frac{(M_{S_i}-M_{h})^2}{4m_\chi^2}}\sqrt{1-\frac{(M_{S_i}+M_{h})^2}{4m_\chi^2}} \hat{y}_{ih}^2,\\
b_{\chi\bar{\chi}\rightarrow t\bar{t}} &=& \frac{N_c y_\chi^2 y_t^2 \epsilon_u^2\sin^2\left(2\theta\right)}{64\pi}\sqrt{1-\frac{m_t^2}{m_\chi^2}}\frac{\left(m_\chi^2-m_t^2\right)\left(M_{S_1}^2-M_{S_2}^2\right)^2}{\left(4m_\chi^2-M_{S_1}^2\right)^2\left(4m_\chi^2-M_{S_2}^2\right)^2}, \eea
where
\bea
\hat{y}_{ij} &=& y_{ij} \quad \left(s\rightarrow 4m_\chi^2\right).
\eea

\subsection{General Relic Density Calculation Formalism}
\label{sec:relicdensityformalism}

We use {\tt micrOMEGAs} \citep{Barducci:2016pcb} to make our parameter scan. In regions away from resonances and thresholds, however, it is much faster to use the expressions from \ref{sec:relicdensitycs} to make an analytical calculation. The relevant formalism is defined in \citep{Gondolo:1990dk,Bertone:2004pz}. All the necessary formulae can be found in Appendix B of \citep{Busoni:2014gta}. Here we cite only the final ones, giving the expression of the relic density and freezeout temperature:

\begin{equation}
e^{x_F} = \frac{\sqrt{\frac{45}{8}} g m_\chi M_{\rm Pl} c(c+2) \sigma v(s)}{2 \pi^3 g_\star^{1/2} \sqrt{x_F}},\label{eq:xf}
\end{equation}

\begin{equation}
\Omega_{\rm DM} h^2  = \frac{g\times 1.04\times 10^9 \, {\rm GeV}^{-1} x_F}{g_\star(T_F)^{1/2}  \, M_{\rm Pl} \,(a + 3 b / x_F)}.
\label{relic-simple}
\end{equation}
where $x_F = m_\chi / T_F$, where $T_F$ is the freezeout temperature, $g$ is the number of degrees of freedom of the DM particle, and is 2 for a Dirac fermion or 1 for a Majorana fermion, $c$ is an $\mathcal{O}(1)$ constant, $g_\star$ is the number of relativistic degrees of freedom, and $M_{\rm Pl} = 1/\sqrt{G_N}$ is the Planck mass.
In our case, when using the wave expansion we will always get the s-wave term $a=0$, thus this formula further simplifies to
\begin{equation}
\Omega_{\rm DM} h^2  = \frac{g\times 2.84\times 10^{-11} \,  x_F^2}{g_\star(T_F)^{1/2} \,(b / {\rm GeV}^{-2})}.
\label{relic-p}
\end{equation}

By solving Eq.~\ref{eq:xf} numerically, one can find a simple fitting function for the value of $x_F$ (and of the relic density) in the case of pure p-wave processes:
\bea
x_F(b,m_\chi) &=& 23.02 + 2.17\log_{10}\left(\frac{b}{10^{-8}\GeV^{-2}} \frac{m_\chi}{100\GeV} \left(\frac{106.75}{g_\star}\right)^{1/2}\right).
\label{eq:xfn}
\eea
One can use Eq.~\ref{relic-p} together with Eq.~\ref{eq:xfn} and the p-wave expansion coefficients listed in Section \ref{sec:relicdensitycs} to calculate the relic density analytically with a good degree of precision, if away from resonances and thresholds.

Purely on dimensional grounds, if the DM particle couples with a strength $y_\chi$, then the thermally averaged cross section for p-wave annihilation is expected to be roughly
\begin{align}
\label{eq:sigmav} \langle \sigma v \rangle \sim \frac{y_\chi^4}{16 \pi^2 m_\chi^2} v^2 + \mathcal{O} (v^4).
\end{align}
Using this reference value for the cross section, plugging it inside Eq.~\ref{relic-p} and requiring the right relic density to be achieved, one would get the relation
\begin{align}
y_\chi \sim \pm 0.047 \sqrt{\frac{m_\chi}{\text{GeV}}}
\end{align}
which is the blue line plotted in Fig.~\ref{fig:scan1} (a). Note that this means the coupling $y_\chi$ will exceed $4\pi$ at $m_\chi \sim  \mathcal{O}(100\TeV)$ \citep{Griest:1989wd}.

\subsection{Sommerfeld Enhancements}
\label{sec:somm}
In the region of the parameter space that we consider, some points can get $\mathcal{O}(1)$ factors from Sommerfeld Enhancement, as we need at least one of $S_{1,2}$ to be lighter than DM. The usual way to account for Sommerfeld Enhancement is to make the wave expansion of the cross section
\bea
\left(\sigma v\right)_{tree}  = \sum_{l=0}^\infty c_l v^{2l} = a + b v^2 +\mathcal{O}(v^4),
\eea
and then multiply each term by the appropriate Sommerfeld Enhancement factor $E_i$,
\bea
\left(\sigma v\right)_{non-pert}  = \sum_{l=0}^\infty c_l E_l v^{2l} = a E_0 + b E_1 v^2 +\mathcal{O}(v^4).
\eea
One can obtain an analytic expression by approximating the Yukawa potential with an Hulth´en potential. The resulting enhancement factors for the first 2 waves are \citep{Feng:2010zp,Cassel:2009wt,Iengo:2009ni,Kahlhoefer:2017umn} 
\bea
E_0&=&\frac{\pi}{a}\frac{\sinh 2\pi a c}{\cosh 2\pi a c - \cos2\pi\sqrt{c-a^2c^2}},\\
E_1&=& \frac{(c-1)^2+4a^2c^2}{1+4a^2c^2}\times E_0,
\eea
where
\begin{align}
a = \frac{v}{2\alpha},  \quad\quad
c = \frac{6\alpha m_\chi}{\pi^2 M_{S_1}}, \quad\quad \textrm{and} 
\quad\quad
\alpha = \frac{y_\chi^2 f_i^2}{4\pi}.
\end{align}

These formulae are not valid where the wave expansion breaks down, however we do not expect Sommerfeld Enhancement to be significant for such points, as being on resonance or threshold usually implies that both scalars are not light enough to make Sommerfeld Enhancement relevant. Indeed we have checked that, for the points for which the expansion breaks down, the would-be Sommerfeld Enhancement factors $E_i$ are negligible for all the points sampled.

\section{2HDM+S -- Gauge Basis to Higgs Basis}
\label{sec:gaugetohiggs}

We considered this model in a previous paper, \cite{Bell:2016ekl}, by instead writing the fields in the gauge basis -- where the two Higgs doublets are $\Phi_1$ and $\Phi_2$. In this appendix we will summarize the relevant equations expressing the physical masses and mixing angles in such a basis, and we will show how to convert the parameters from the gauge basis to the so called Higgs basis, where one of the doublets has no vev; $\langle \Phi_H \rangle =0$.
The scalar potential considered has an additional Higgs doublet and a singlet scalar $S$:
\begin{equation}
V(\Phi_1,\Phi_2,S) = V_{\mathsc{2hdm}}(\Phi_1,\Phi_2) + V_S(S) + V_{S\mathsc{2hdm}}(\Phi_1,\Phi_2,S), 
\end{equation}
where\nobreak
\bea
V_{\mathsc{2hdm}}(\Phi_1,\Phi_2) &=& M_{11}^2 \Phi_1^\dagger \Phi_1 + M_{22}^2 \Phi_2^\dagger \Phi_2 +  (M_{12}^2 \Phi_2^\dagger \Phi_1 + h.c.) + \frac{\lambda_1}{2} (\Phi_1^\dagger \Phi_1)^2 + \frac{\lambda_2}{2} (\Phi_2^\dagger \Phi_2)^2 
\nonumber \\
&+&\lambda_3 (\Phi_1^\dagger \Phi_1)(\Phi_2^\dagger \Phi_2) + \lambda_4 (\Phi_2^\dagger \Phi_1)(\Phi_1^\dagger \Phi_2) + \frac{1}{2}\left(\lambda_5 (\Phi_2^\dagger \Phi_1)^2 + h.c.\right), \\
V_S(S) &=& \frac{1}{2} M_{SS}^2 S^2 + \frac{1}{4} \lambda_S S^4,
\\
V_{S\mathsc{2hdm}}(\Phi_1,\Phi_2,S) &=& 
\frac{\lambda_{11S}}{2}(\Phi_1^\dagger \Phi_1)S^2 +  \frac{\lambda_{22S}}{2}(\Phi_2^\dagger \Phi_2)S^2 + \frac{1}{2}(\lambda_{12S} \Phi_2^\dagger \Phi_1 S^2 + h.c.).
\eea

The fields are expanded around their vevs, and the minima conditions are imposed; eliminating $M_{11}^2$, $M_{12}^2$, and $M_{S}^2$. We define
\begin{equation}
\tan\beta =  \frac{v_2}{v_1}, \;\;\;\; {\textrm{with}} \;\;\;\; 
v_1^2+v_2^2 = v^2,
\end{equation}
and impose the alignment condition
\begin{align} 
\lambda_3 &=
  \frac{1}{2}\left(\lambda_1+\lambda_2-2\lambda_4-2\lambda_5
  +\left(\lambda_1-\lambda_2\right)\sec
  2\beta\right),
\label{eq:alignment}\\ 
\lambda_{11S} &=
  -\tan\beta\left(2\lambda_{12 S}+\lambda_{22S} \tan\beta\right).\label{eq:halign}
\end{align} 
Eq.~\eqref{eq:alignment} is always satisfied in the presence of a CP2 symmetry \citep{Dev:2014yca} that imposes
\begin{equation}
\lambda_1=\lambda_2=\lambda_3+\lambda_4+\lambda_5.
\end{equation}
This condition will be referred to as the CP2 condition. Note that this first condition also sets $\hat{\lambda}_6=\hat{\lambda}_7=0$ in the Higgs basis, provided that $\lambda_6=\lambda_7=0$ in the gauge basis, given Eq.~\eqref{eq:alignment}, \eqref{lam6} and \eqref{lam7}.
\\

Diagonalizing the mass matrices, one can find the mass spectrum and mixing angle of the scalars:
\begin{eqnarray}
M_{H^+}^2 &=& \sec ^2\beta \left(M_{22}^2+\frac{1}{2} \lambda_{22S} v_S^2\right)+\frac{v^2}{2} \left(\lambda_2 \tan ^2\beta+\lambda_{3}\right),\\
M_A^2 &=& \sec ^2\beta \left(M_{22}^2+\frac{1}{2} \lambda_{22S} v_S^2\right)+ \frac{v^2}{2} \left(\lambda_2 \tan ^2\beta+ 
   \lambda_3+\lambda_4-\lambda_5 \right),\\
M_{S_{1,2}}^2 &=&
\frac{1}{2}\left(M_A^2+\lambda_5 v^2+ \left(\lambda_2
v^2-M_h^2\right)\tan^2 \beta\right)\left(1\pm \frac{1}{\cos
  2\theta}\right) + \lambda_S v_S^2\left(1\mp \frac{1}{\cos
  2\theta}\right), \\
\tan2\theta &=& \frac{4 v \cos
  ^2\beta v_S \left(\tan \beta \lambda_{22S}+\lambda _{12
    S}\right)}{\cos 2 \beta \left(M_A^2+M_h^2-2 \lambda _S
  v_S^2-\lambda _2 v^2+\lambda_5 v^2\right)+M_A^2-M_h^2-2 \lambda _S
  v_S^2+\lambda _2 v^2+\lambda _5 v^2}, \quad
\end{eqnarray}
and one can rewrite the Lagrangian in terms of the mass eigenstates
\bea
\Phi_h &=& \cos\beta \Phi_1 + \sin\beta \Phi_2 = \left(
\begin{array}{cc}
 G^+ \\
\frac{v + h + i G^0}{\sqrt{2}} \\
\end{array}
\right),\\
\Phi_H &=& -\sin\beta \Phi_1 + \cos\beta \Phi_2 = \left(
\begin{array}{cc}
 H^+ \\
\frac{ H + i A}{\sqrt{2}} \\
\end{array}
\right),\\
H &=& \cos\theta S_1 - \sin\theta S_2,\\
S &=& v_S + \sin\theta S_1 + \cos\theta S_2. 
\eea
It is convenient to rewrite the same potential in the Higgs basis, where $\langle \Phi_H \rangle =0$. One gets 
\begin{equation}
\hat{V}(\Phi_h,\Phi_H,S) = \hat{V}_{\mathsc{2hdm}}(\Phi_h,\Phi_H) + \hat{V}_S(S) + \hat{V}_{S\mathsc{2hdm}}(\Phi_h,\Phi_H,S), 
\end{equation}
where\nobreak
\bea
\hat{V}_{\mathsc{2hdm}}(\Phi_h,\Phi_H) &=& \hat{M}_{hh}^2 \Phi_h^\dagger \Phi_h + \hat{M}_{HH}^2 \Phi_H^\dagger \Phi_H +  (\hat{M}_{hH}^2 \Phi_H^\dagger \Phi_h + h.c.) + \frac{\hat{\lambda}_h}{2} (\Phi_h^\dagger \Phi_h)^2 + \frac{\hat{\lambda}_H}{2} (\Phi_H^\dagger \Phi_H)^2 
\nonumber \\
&+&\hat{\lambda}_3 (\Phi_h^\dagger \Phi_h)(\Phi_H^\dagger \Phi_H) + \hat{\lambda}_4 (\Phi_H^\dagger \Phi_h)(\Phi_h^\dagger \Phi_H)  
\nonumber \\
&+& \frac{1}{2}\left(\hat{\lambda}_5 (\Phi_H^\dagger \Phi_h)^2 + \hat{\lambda}_6 (\Phi_h^\dagger \Phi_h)(\Phi_H^\dagger \Phi_h) + \hat{\lambda}_7 (\Phi_H^\dagger \Phi_H)(\Phi_h^\dagger \Phi_H) + h.c.\right),
\\
\hat{V}_S(S) &=& \frac{1}{2} \hat{M}_{SS}^2 S^2 + \frac{1}{4} \hat{\lambda}_S S^4,
\\
\hat{V}_{S\mathsc{2hdm}}(\Phi_h,\Phi_H,S) &=& 
 \frac{\hat{\lambda}_{HHS}}{2}(\Phi_H^\dagger \Phi_H)S^2 + \frac{1}{2}(\hat{\lambda}_{hHS} \Phi_H^\dagger \Phi_h S^2 + h.c.).
\eea
Note that we have omitted a term of the form $\Phi_h^\dagger \Phi_h S^2$, as the coefficient $\hat{\lambda}_{hhS}$ for this term is set to be zero by the alignment condition \ref{eq:halign}.
The new couplings are related to those in the previous basis by
\bea
\hat{\lambda}_h &=& \cos^4\beta \lambda_1 + \sin^4\beta \lambda_2 + 2\sin^2\beta\cos^2\beta \left(\lambda_3 +\lambda_4+\lambda_5\right) \stackrel{CP2}{=}\lambda_1, \\
\hat{\lambda}_H &=& \sin^4\beta \lambda_1 + \cos^4\beta \lambda_2 + 2\sin^2\beta\cos^2\beta \left(\lambda_3 +\lambda_4+\lambda_5\right) \stackrel{CP2}{=}\lambda_1, \\
\hat{\lambda}_3 &=& \sin^2\beta\cos^2\beta \left(\lambda_1+\lambda_2-2\lambda_4-2\lambda_5\right) + \left(\cos^4\beta+\sin^4\beta\right)\lambda_3  \stackrel{CP2}{=}\lambda_1 - \lambda_4 - \lambda_5, \\
\hat{\lambda}_4 &=& \sin^2\beta\cos^2\beta \left(\lambda_1+\lambda_2-2\lambda_3-2\lambda_5\right) + \left(\cos^4\beta+\sin^4\beta\right)\lambda_4  \stackrel{CP2}{=}\lambda_4, \\
\hat{\lambda}_5 &=& \sin^2\beta\cos^2\beta \left(\lambda_1+\lambda_2-2\lambda_3-2\lambda_4\right) + \left(\cos^4\beta+\sin^4\beta\right)\lambda_5  \stackrel{CP2}{=}\lambda_5, \\
\label{lam6} \hat{\lambda}_6 &=& -\sin\beta\cos\beta \left(\lambda_1 - \lambda_2 + \cos 2 \beta (\lambda_1 + \lambda_2 - 2\lambda_3 - 2\lambda_4 - 2\lambda_5)\right) \stackrel{CP2}{=}0, \\
\label{lam7} \hat{\lambda}_7 &=& -\sin\beta\cos\beta \left(\lambda_1 - \lambda_2 - \cos 2 \beta (\lambda_1 + \lambda_2 - 2\lambda_3 - 2\lambda_4 - 2\lambda_5)\right) \stackrel{CP2}{=}0, \\
\hat{\lambda}_s &=& \lambda_s, \\
\hat{\lambda}_{hHS} &=& \lambda_{12S} \cos 2\beta  + (\lambda_{22S} - \lambda_{11S}) \cos \beta \sin \beta \stackrel{align}{=} \lambda_{12S} +\tan\beta \lambda_{22S}\\
\hat{\lambda}_{HHS} &=& \lambda_{22S} \cos^2 \beta - 2 \lambda_{12S} \cos \beta \sin \beta + \lambda_{11S} \sin^2 \beta \stackrel{align}{=} \lambda_{22S}\left(1-\tan^2\beta\right) -2\lambda_{12S} \tan\beta. \quad \quad 
\eea
Note that, thanks to the CP2 symmetry, when rewriting the model in the Higgs basis (or any other basis related to the old one by a rotation of the two doublets), the values of the 2HDM couplings do not change, and do not depend on the value of $\tan\beta$.


\newpage

\label{Bibliography}

\lhead{\emph{Bibliography}} 

\bibliography{Bibliography} 

\end{document}